\documentclass[preprint,12pt, authoryear]{elsarticle}

\usepackage{amssymb}
\usepackage{float, tabularx, makecell, multirow, balance, booktabs,soul}

\usepackage[most]{tcolorbox}
\usepackage{hyperref}
\usepackage{adjustbox}
\usepackage{natbib}
\usepackage{lscape}
\usepackage{amsmath}
\usepackage{xcolor}
\usepackage{tikz}
\usepackage{longtable} 
\newcommand*\circled[1]{\tikz[baseline=(char.base)]{
            \node[shape=circle,draw,inner sep=1pt] (char) {#1};}}

\tcbset{textmarker/.style={%
        enhanced,
        parbox=false,boxrule=0mm,boxsep=0mm,arc=0mm,
        outer arc=0mm,left=1.5mm,right=1mm,top=5pt,bottom=5pt,
        toptitle=1mm,bottomtitle=1mm,oversize}}

\newtcolorbox{noteBox}{
    breakable,
    enhanced,
    textmarker,
    borderline west={3pt}{0pt}{gray},
    before skip=10pt,
    after skip=5pt,
    colback=gray!10!white
}

\usepackage{url}
\usepackage{colortbl}
\usepackage{xcolor}
\usepackage{microtype}
\journal{}

\begin{document}

\begin{frontmatter}



\title{Investigating Software Developers' Challenges for \\ Android Permissions in Stack Overflow}


\author{Sahrima Jannat Oishwee, Natalia Stakhanova, Zadia Codabux}

\affiliation{organization={Department of Computer Science, University of Saskatchewan},
            city={Saskatoon},
            state={Saskatchewan},
            country={Canada}}

\begin{abstract}
The Android permission system is a set of controls to regulate access to sensitive data and platform resources (e.g., camera). The fast evolving nature of Android permissions, coupled with inadequate documentation, results in numerous challenges for third-party developers. This study investigates the permission-related challenges developers face and the solutions provided to resolve them on the crowdsourcing platform Stack Overflow. We conducted qualitative and quantitative analyses on 3,327 permission-related questions and 3,271 corresponding answers. Our study found that most questions are related to non-evolving SDK permissions that remain constant across various Android versions, which emphasizes the lack of documentation. We classify developers' challenges into several categories: ~\textit{Documentation-Related, Problems with Dependencies, Debugging, Conceptual Understanding, and
Implementation Issues}. We further divided these categories into 12 subcategories, nine sub-subcategories, and nine sub-sub-subcategories. Our analysis shows that developers infrequently identify the restriction type or protection level of permissions, and when they do, their descriptions often contradict Google's official documentation. Our study indicates the need for clear, consistent documentation to guide the use of permissions and reduce developer misunderstanding leading to potential misuse of Android permission. These insights from this study can inform strategies and guidelines for permission issues. Future studies should explore the effectiveness of Stack Overflow solutions to form best practices and develop tools to address these problems.

\end{abstract}

\begin{graphicalabstract}
\end{graphicalabstract}

\begin{highlights}
\item Categorizing developers' challenges related to Android permissions 
\item Categorizing solutions to developers' challenges by the Stack Overflow community 
\item Identifying error-prone restriction types \& protection levels of permissions 
\item Comparing community-provided information with Google's official documentation 
\end{highlights}


\end{frontmatter}


\section{Introduction}

Android has quickly become one of the most popular mobile Operating Systems (OS), providing a flexible and versatile environment for building applications (apps). It has an open-source nature and flexible platform, which allow developers to create diverse apps, often with extensive access to phone resources. To regulate access to sensitive resources (e.g., camera) and data, Android implemented a system of safeguards called permissions\footnote{\url{https://developer.android.com/guide/topics/permissions/overview\#system-components}}. An app with access to restricted resources must request the relevant permissions, which can then be granted by the Android OS. 
Over the years, the permission system has evolved from almost unrestricted access to advanced phone functionalities to a fine-grained and strictly regulated approach. Beginning with Android 9, permission restrictions on interfaces limited to third-party developers (referred to as non-Software Development Kit (SDK) interfaces) were introduced, further restricting access to system parts of the Android platform. These permissions for non-SDK interfaces are not officially documented or supported by the Android SDK. The restricted permissions are typically used by system apps and thus provide access to special features or resources on an Android device that are not intended for general app development. Numerous studies reported limitations of the Android permission system and its use for SDK and non-SDK interfaces, including security challenges (\cite{he2022systematic}) and development problems (\cite{wei2016taming, huang2018understanding, vidas2011curbing}). 

Developers can utilize a wide range of resources, including official documentation, introductory guides, Q\&A forums, mailing lists, and blog posts, to acquire knowledge about SDK permissions. However, it is not always apparent to what extent these resources comprehensively address API and the associated permission topics (\cite{ajam2020api}).~\cite{Tuncay20} showed that users have a vague understanding of which permissions to grant or deny to apps. Developers have also raised concerns indicating, among other challenges, a struggle to keep up with the fast-evolving Android permission system and APIs,  often resulting in security risk and compatibility issues (\cite{wei2016taming,taylorevolution2017, huang2018understanding, mahmudAndroid2021}). ~\cite{Krutz17} showed that less experienced developers are likely to make permission-related changes early in the app's commit lifetime. 


Even with an advanced safeguards system, documentation has not been given much importance by Android. The official documentation for Android permissions and their restrictions has historically been inadequate (e.g.,~\cite{felt2011android}), lacking details of the requirements, dependencies, and interfaces they protect.~\cite{Dawoud21} found that incomplete documentation for sensitive Android APIs results in permission misconfigurations. 

Android has been restricting non-SDK permissions to enhance user privacy and security; hence, non-SDK interfaces and their corresponding permissions have not been officially documented. Yet, several studies showed that permissions are silently reclassified over time, transitioning from non-SDK to SDK categories and vice versa (\cite{li2016accessing}). The evolution of permission system for SDK and non-SDK interfaces remains vaguely defined, contributing to the ambiguity in permission interpretations and leading to uncertainty and errors in app implementation.

In this study, we focus on common challenges encountered by Android developers when using the Android permission system. We investigate the concerns raised by developers and explore solutions provided by the Android community in response to these concerns. Our study is based on discussions posted on the Stack Overflow platform\footnote{\url{https://stackoverflow.com/}}, one of the largest forums that attract an actively engaged community of developers who offer comprehensive technical information, practical solutions, and best practices. For our analysis, we collected 3,327 questions encompassing 185 Android permissions posed by Android developers and 3,271 corresponding responses. 

Through our analysis, we derived a fine-grained categorization of challenges faced by developers and the corresponding categorization of responses. We observed that developers face challenges that fall into several categories: \textit{Documentation-Related, Problems with Dependencies, Debugging, Conceptual Understanding, and Implementation Issues}. The community-provided responses are categorized into \textit{Conceptual Understanding, References to Documentation, Debugging Steps, Implementation, and Advice for Configuration Changes.}

Our results show that most developers' questions pertain to officially documented SDK permissions, which have consistently maintained their classification and remained stable across different Android versions. This finding emphasizes the insufficiency of official documentation, emphasizing the necessity for consistent and up-to-date resources to address the challenges effectively.

Our analysis also shows that developers are facing inadequate support from the community. 32.8\% of questions remain unanswered, and the rate of accepted answers is low, standing at 30.2\%. This lack of assistance could be similarly attributed to Google's limited documentation.



To summarize, our study has the following contributions:

\begin{itemize}
\item We have categorized permission-related challenges encountered by developers and the corresponding solutions provided by the community. Our categorization provides insights into Android's permission usage difficulties and highlights the impact of insufficient documentation, and it's evolving nature. This categorization aims to identify knowledge gaps and potential solutions, consequently facilitating research in this area and guiding tool development to address the existing challenges.

\item We summarized the permission restriction types and protection levels that pose challenges to developers and evaluated the accuracy of community-provided information on these aspects. These insights can guide Google to prioritize certain permissions when improving its documentation.

\item We assessed the accuracy of information shared within the developer community regarding permission types in relation to the official documentation, determining whether the community effectively conveys correct information about permission types. This information will highlight the among developers regarding permission restriction types and protection levels.

\item We investigated the types of solutions, rate of provided answers, accepted answer rate, and up and downvoted answers. This finding indicates the need for a support tool for developers to mitigate permission-related issues. 

\item To facilitate research in this area, we provide a replication package\footnote{\url{https://doi.org/10.5281/zenodo.8226832}} consisting of the permission lists and restriction types, selected permissions, dataset of permission-related posts on Stack Overflow, and descriptions of categories.

\end{itemize}

\section{Background}

One of Android's fundamental access control mechanisms to manage access to sensitive phone resources and data is the \textit{Android permission system}. Third-party developers must declare the necessary permissions in the app's AndroidManifest file containing essential metadata about the app, including various configuration settings, required permissions, and declared app's activities, services, and functionalities. Android OS can then determine whether to grant or decline the requested permissions.

The Android permission system has significantly evolved over time. The early version of Android allowed granting permissions to apps during  installation\footnote{\url{https://web.archive.org/web/20130822155513/https://developer.android.com/training/articles/security-tips.html}}. Starting from Android 6, permissions considered high risk were granted at the runtime of the app. With Android 9 and onwards, Android further restricted the permission system by introducing safeguards on non-SDK interfaces to regulate access to the Android platform\footnote{\url{https://developer.android.com/guide/app-compatibility/restrictions-non-sdk-interfaces}}. 
These restrictions were listed in the \textit{restriction lists} that have since been officially published for each major version of Android. 

\paragraph{\textbf{Permission categorization}} Android official documentation categorizes  permissions according to their protection level that characterizes the potential risk implied in the permission\footnote{\url{https://developer.android.com/guide/topics/manifest/permission-element}} as follows: 
(1) \textit{Normal permissions} are minimal risk permissions, (2) \textit{Dangerous permissions} pose a high level of risk (e.g., access to phone camera or microphone),   (3) \textit{Signature permissions} is granted when the app requesting it has the same certificate as the app that defined the permission,
(4)\textit{SignatureOrSystem}(Signature$\big|$Privileged permissions)  are special types of permissions that are granted exclusively to system apps or those signed with the same certificate as the app declaring the permission. 

Restrictions lists contain permissions with the corresponding interfaces and restriction categories. Officially, the restriction lists include the following categories\footnotemark[5]:

(1) \textit{SDK (whitelist)} permissions that are formally recognized and explicitly explained in the official documentation. The Software Development Kits (SDKs), which have their documentation included in the package index, are commonly referred to as the public API.

(2) \textit{Blocklist (blacklist)} permissions are forbidden from using for third-party developer,

(3) \textit{Conditionally blocked (greylist-max-x)} permits the use of certain permissions and corresponding interfaces for applications that target an API level up to `x', but restricts access for others, and 

(4) \textit{Unsupported (greylist)} permissions that, although they were not limited at the time of launch, are not included in the documentation, which leaves them susceptible to changes without any prior notice.

Our analysis revealed that the restriction lists include other categories (e.g., system-API, test-API, removed, lo-prio) that are not officially documented.

\section{Related Work} 
A broad analysis has been done of the Android permissions system since its introduction in 2008. The analysis from the study of ~\cite{almomani2020comprehensive} shows a seven times increase in permission numbers since Android's initial release, suggesting significant growth. A case study examining the last five years of top Android apps revealed an evolution in the permissions system and related security issues, with some apps showing a 73.3\% increase in permission usage by 2020 (\cite{almomani2020comprehensive}). 
This increase, however, is typically due to apps requesting unnecessary permissions, a phenomenon known as overprivileged apps (\cite{felt2011android}). For example, ~\cite{au2012pscout} found that about 22.0\% of non-system permissions are unnecessary.~\cite{taylorevolution2017} analyzed 30K Android applications on Google Play for two years (Oct 2014 to Sep 2016) and found that many applications ask for additional dangerous permissions, and Android updates do not make them safer. Free and popular apps are more likely to ask for additional permissions, and updates can increase the number of vulnerabilities. 
To understand the misuse of permissions, ~\cite{stevens2013asking} analyzed 10,000 free Android apps. They built a statistical model and found that the popularity of permission is strongly associated with its misuse. They also investigated   Stack Overflow data and found that more widely used permissions get mentioned more often in questions on Stack Overflow.~\cite{vidas2011curbing} mentioned that the responsibility for determining which permissions to request rests solely on the developers, leading to instances where certain applications solicit permissions they do not actually need.

To address this problem, several solutions aiming for automatic assessment of the requested permissions and their need for the app's functionality were proposed, e.g., 
Arcade (\cite{aafer2018precise}), Dynamo (\cite{Dawoud21}), PmDroid (\cite{gao2015}), IntentChecker (\cite{he2016detecting}), StaDART (\cite{ahmad2020stadart}). 
~\cite{Dawoud21} presented Dynamo, a dynamic testing tool designed to analyze Android's application framework security policy and validate static analysis tools' results. Their consistency analysis of the permission revealed that five sensitive APIs are unprotected and 65 APIs with permission misconfigurations. In addition to that, they also found incomplete documentation for 66 APIs and imprecise documentation for 9 APIs, which they reported to Google for corrective actions.~\cite{ahmad2020stadart} introduced StaDART, which integrates static and dynamic analysis to detect hidden malware activities in Android apps for API used in the app. It is automated and scalable, merging with a triggering solution, DroidBot (\cite{li2017droidbot}). Evaluation across 2000 real-world apps demonstrated StaDART's efficacy in uncovering suspicious behavior missed by static analysis tools. 

To reduce the misuse of permissions by users ~\cite{wei2015claim} explored how an app's description relates to the permissions it requests. They developed a model to help app users predict what permissions the app should rightfully ask for based on its description. ~\cite{anwer2014chiromancer} introduced a tool called Chiromancer, designed to manage an app's permissions with high precision and regulate the app's resource usage. ~\cite{li2018characterising} built a tool, CDA, to analyze deprecated APIs in the Android framework, assessing their prevalence, documentation, and replacement procedures. The results identify three bugs tied to deprecated APIs.

~\cite{he2022systematic} provided a systematic examination of vulnerabilities resulting from hidden API exploitation and evaluated the effectiveness of Google's countermeasures. They introduced ServiceAudit, a static analysis tool developed to automatically detect inconsistent security enforcement between service helper classes and hidden service APIs. ServiceAudit found 112 vulnerabilities in Android 6 with higher precision than previous methods and revealed more than 25 hidden APIs with inconsistent protections in Android 11 and 12, though few led to severe security issues. ~\cite{felt2011permission} discussed the concept of permission redelegation in Android OS, which undermined user approval requirements and can lead to potential security vulnerabilities. To address this issue, the authors introduced Inter Process Communication (IPC) inspection, a new mechanism that limits an app's permissions when it interacts with a less privileged one, demonstrating its effectiveness in preventing real-world attacks on Android apps. To facilitate analysis of the correlation between permission and code smells, developers conducted permission-related tasks, ~\cite{Scoccia19}, collected a dataset of permission-based changes and permission-related issues from 2,002 apps from 10,601 unique commits from F-Droid. To enhance the Android permission system, ~\cite{barrera2010methodology} analyzed 1,100 Android applications and suggested areas of potential enhancement for the Android permission model, aiming to boost its expressiveness without adding to the total number of permissions or the system's overall complexity. Android permission protocol safeguards against unauthorized access, ensuring each application can only access the components for which it has been granted permissions and nothing beyond. ~\cite{bagheri2018formal} analyzed the Android permission protocol for applications. They described an automatic analysis that identifies potential flaws in the protocol and demonstrated through real-world apps that these flaws can have severe security implications. ~\cite{wei2016taming} conducted an empirical study on 191 issues collected from open-source Android apps. This study revealed the common types, symptoms, and fixes for fragmentation-induced compatibility issues. When the Android app control flows are guided by callback APIs, compatibility issues can arise unexpectedly as these APIs evolve with the Android framework. To address this, ~\cite{huang2018understanding} analyzed Android documentation and studied 100 real-world callback compatibility issues, focusing on their origins in callback API evolution. They created a graph-based model capturing control flow inconsistencies caused by API evolution and developed Cider, a static analysis technique to detect these issues. Evaluating Cider on 20 open-source Android apps demonstrated its effectiveness, detecting 13 new callback compatibility issues, 12 confirmed and nine fixed. To mitigate compatibility issues, ~\cite{scalabrino2019data} proposed ACRYL, a data-driven methodology that learns from modifications made in other applications due to API changes. This approach not only identifies potential compatibility problems but also provides recommendations for resolution.~\cite{mahmudAndroid2021} mentioned that Android apps developed using a SDK and an API frequently face compatibility issues due to the evolution of APIs and SDKs. These issues can result in unexpected behaviors, such as runtime crashes, which hamper user experience. To address this issue, they introduced ACID, a novel approach to detecting compatibility issues induced by API evolution. ~\cite{backes2016demystifying} provided a comprehensive analysis focusing on the permission-based privilege model of the Android application framework. They presented new insights into analyzing the Android application framework from a security perspective and the framework's security challenges. ~\cite{li2016accessing} analyzed the permissions and identified inconsistent security policy enforcement within the Android framework, which can pose security and privacy risks to users. ~\cite{xia2020android} focused on the Android developers' reactions to evolution-induced API compatibility issues. Their research showed that developers do not want to provide alternative implementations for incompatible API invocations. ~\cite{ajam2020api} studied 400 API-related threads on Stack Overflow and found that the community provided helpful support for API usage, debugging, and security. ~\cite{villanes2017software} analyzed Android testing topics on Stack Overflow using the topic modeling with Linear Discriminant Analysis (LDA) algorithm and found that functional testing, unit testing, and testing tools were frequently discussed. ~\cite{linares2014api} investigated a relationship between API changes and developers' reactions. They studied 213836 Android-related questions on Stack Overflow and found that Android developers have more questions when the API behavior is modified. ~\cite{Krutz17} investigated who makes crucial permission decisions in Android apps and when in the development lifecycle these choices occur, analyzing 1402 Android version control repositories with over 331318 commits. Their findings reveal that experienced developers are more likely to make permission-related changes and that these permissions are usually added early in the app's commit lifetime, but removal extends throughout the lifecycle, and changes are usually reverted by more experienced developers.

\textbf{Summary} The previous research focused on compatibility issues, misuse of permissions, vulnerabilities related to permissions, and tools for resolving those challenges. Here only a few studies focused on the developers' perspective on permission-related issues. Though ~\cite{linares2014api} mentioned that developers have more questions when the API behavior is modified, they did not investigate what type of questions developers ask or the challenges developer face regarding permissions. Moreover, it is crucial to understand the challenges of permissions faced by developers to mitigate the misuse, and security risks in apps. Our study focuses on the permission-related developers' challenges and solutions provided by the community to bridge the knowledge gap. This will also provide the researcher insight into the Android permission-related challenges that need to be mitigated to reduce the misuse of Android permissions.

\section{Methodology}

\subsection{Research Questions}
The goal of this study is to identify the common challenges encountered by Android developers in relation to the use of permissions and examine the solutions provided by the developers-community to tackle these challenges. To achieve our goal, we posed four research questions:

\textbf{RQ1: What are the common challenges developers encounter while using Android permissions?}
    
This research question investigates common challenges developers encounter with permissions, focusing on complexities in adapting to updates of permission systems, maintaining permission-related compatibility issues, and ensuring security across various devices and Android versions. 

\textbf{RQ2: What are the common solutions provided by the community to address the permission-related challenges? }
    
Through this research question, we aim to identify the prevalent solutions proposed by the community to tackle permission-related issues. In addition to that, we investigate the impact of developers' (Stack Overflow users) profiles on provided solutions.

\textbf{RQ3: What are the specific restriction types and permission protection levels that cause the most issues for developers?}

The primary aim of this research question is to investigate the types (in terms of restriction types and protection levels) of permissions that are more prone to cause issues for developers. With the significant transformation of the Android permission system over time, this inquiry also intends to understand how this progression influences the issues developers encounter.

\textbf{RQ4: Does the community provide the correct information about the permissions types that aligns with the official documentation? }
    
With this research question, we are investigating whether the Stack Overflow community accurately conveys information regarding the types (in terms of restriction types and protection levels) of permissions.



\begin{figure}[ht]
    \centering
    \includegraphics[width= 10 cm]{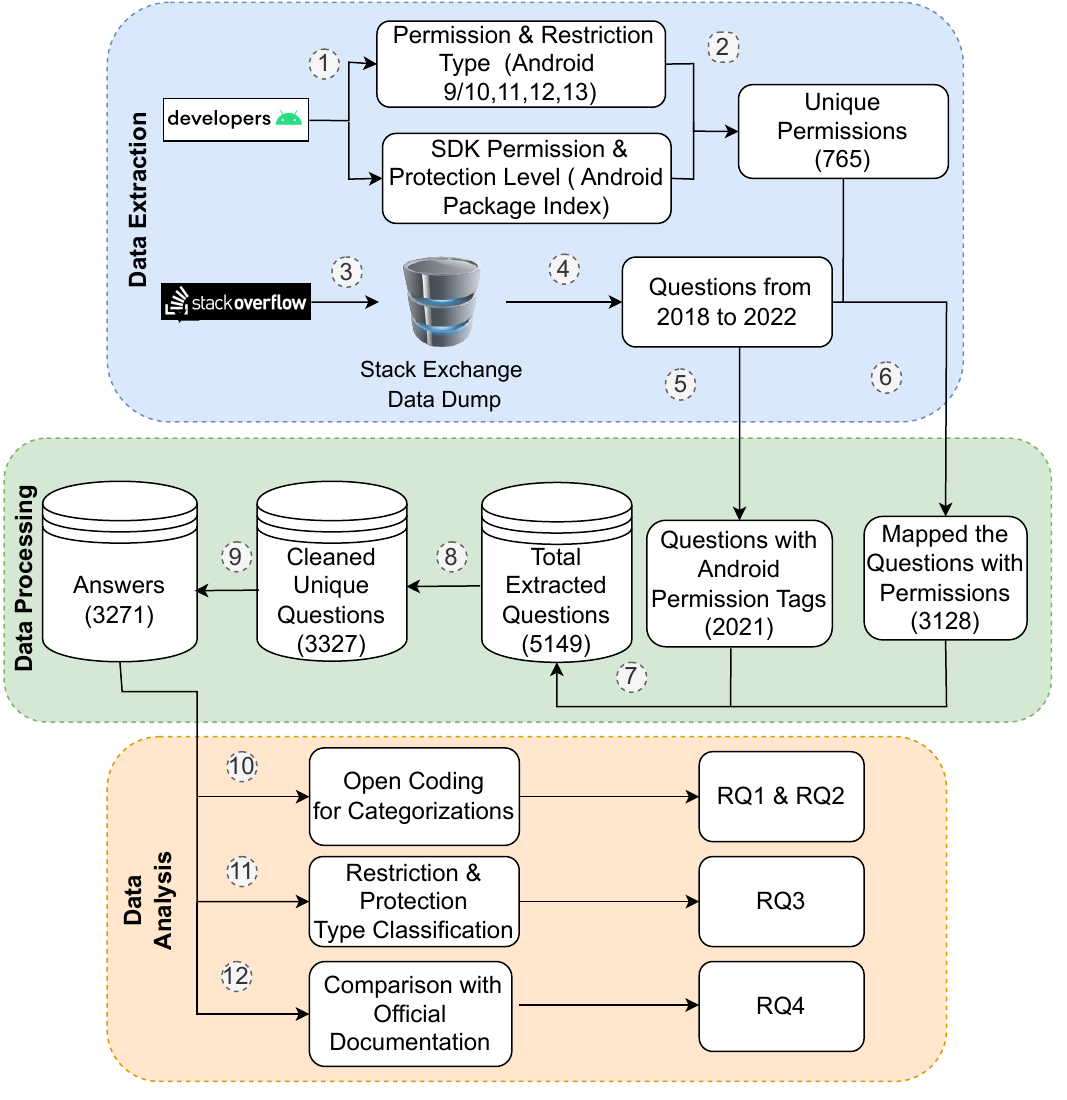}
    \caption{Study Design for Investigating the Android Permission-Related Challenges}
    \label{fig:studydesign}
\end{figure}

\subsection{Study Design}

The methodology, which includes three major components - data extraction, processing, and analysis, is depicted in Figure~\ref{fig:studydesign}.

In step \circled{1}, we parsed restriction lists published by Google for Android 9/10, 11, 12, and 13 and extracted permissions and their corresponding restriction types\footnotemark[5]. For SDK permissions, we complemented this information with the protection levels extracted from the official documentation\footnote{\url{https://developer.android.com/reference/android/Manifest.permission}}. In step \circled{2}, we compiled a unified list of unique permissions and found 765 unique permissions till Android 13  (summarized in Table ~\ref{table:permissionsummary}). 
In step \circled{3}, we extracted Stack Overflow posts from the archive called Stack Exchange data dump\footnote{\url{https://archive.org/details/stackexchange}}. Next, in step \circled{4}, we filtered out the posts from August 2018  to October 2022. This period was defined by the appearance of the restriction lists, i.e., the first list was published for Android 9 and released in August 2018. In step \circled{5}, we extracted all the questions labeled with Android permission-related tags (such as Android, permissions, Android-permissions). This produced 2,021 Stack Overflow posts. Based on our manual verification, we observed that certain permission-related questions were not properly tagged, indicating that this step might have overlooked some relevant posts. 

To address this issue, in step \circled{6}, we mapped the 765 permissions to the titles and content of Stack Overflow questions. As a result, we identified 3,128 posts that contained Android permission-related questions. Combining both sets of 2,021 and 3,128 posts produced  5,149 Android permission-related questions in step \circled{7}. Next, we manually verified and filtered irrelevant and duplicate questions (based on post ID) in step \circled{8}. This resulted in a final set of 3,327 questions. In step \circled{9}, we extracted the 3271 corresponding answers.

Our data analysis incorporated two steps. In step
\circled{10}, we performed qualitative analysis for categorization with question and answer pairs. In the next step \circled{11}, we classified permissions contained within the collected questions. Finally, in step \circled{12}, we compared the community-provided information with Google's official documentation.

\begin{table}[htbp]
\centering
\scalebox{.7}{
\begin{tabular}{lrrrr} 
\hline
Source & ~ Android 9/10 & ~ ~ Android 11 & ~ ~ Android 12 & ~ ~ Android 13 \\ 
\hline
\# Permission & 533 & 589 & 690 & 765 \\
\hline
\end{tabular}}
\caption{The Summary of Permissions Extracted from the Official Restrictions Lists}
\label{table:permissionsummary} 
\end{table}

\subsection{ Data Analysis}

\begin{landscape}
\begin{figure}[htbp]
    \centering
    \includegraphics[width=1.3\textwidth]{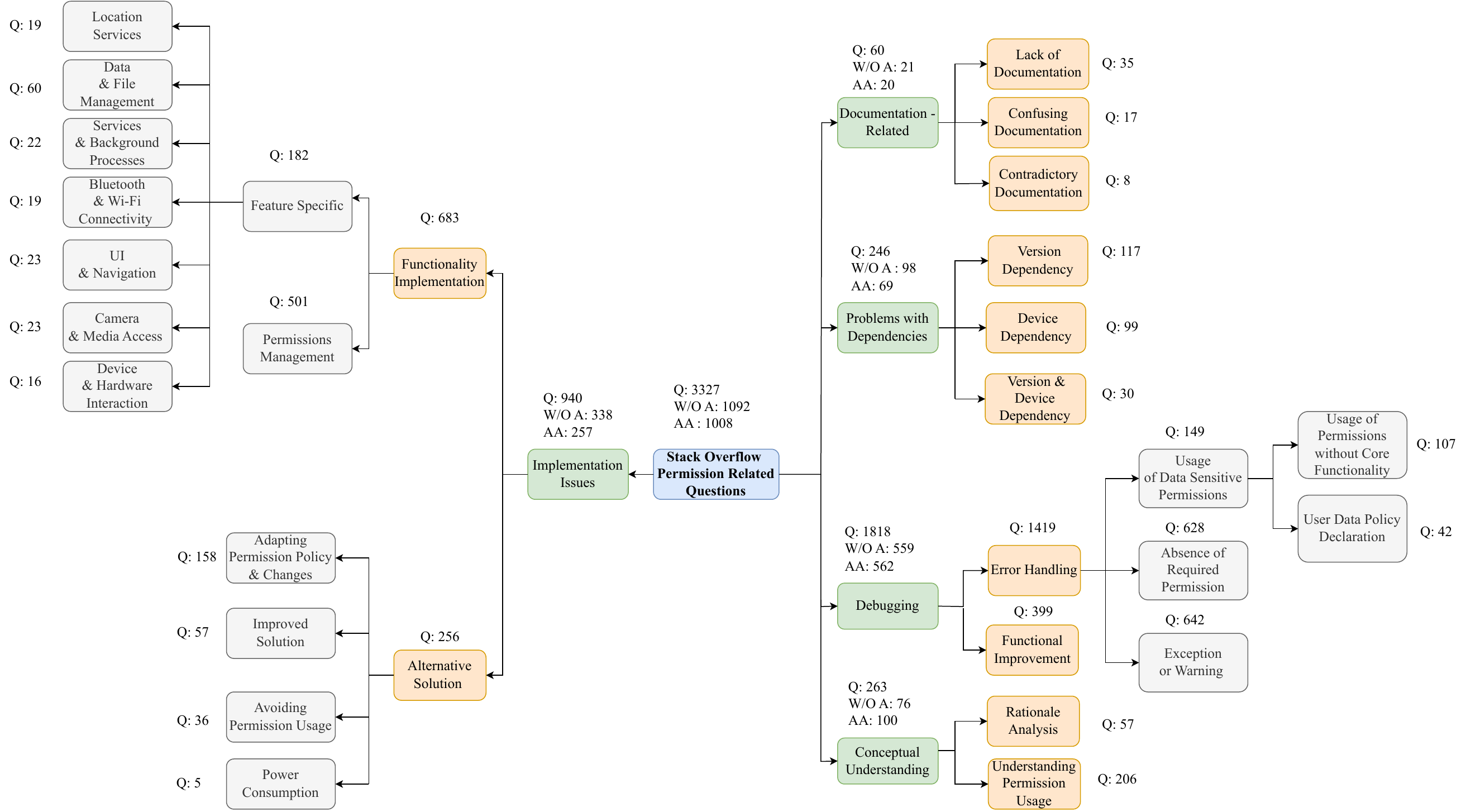}
    \caption{Categorization of Questions Corresponding to Challenges (Q = Number of Question, W/OA = Without Answer, AA = Accepted Answer)}
    \label{fig:classification_questions}
\end{figure}
\end{landscape}

To address RQ1, the 3,327 questions were analyzed in two steps. First, we performed open coding to identify the main categories and patterns. Open coding is a qualitative research method used to analyze data in an exploratory and iterative way (\cite{glaser2016open}). Second, based on the findings from the open coding, we categorized the challenges faced by developers. One author performed open coding on the 3,327 questions. They went through the dataset, identified patterns, and assigned codes to represent the categories that emerged from the data. To ensure accuracy and mitigate the annotation bias, another author independently assessed each question to determine their category. Then, we used Cohen’s Kappa~(\cite{cohen1960coefficient}) to measure inter-rater reliability. The result showed strong agreement among the raters (Cohen’s Kappa=0.87). All disagreements among the raters were resolved by discussing the questions they marked differently.
Any discrepancies or differences in coding were discussed and resolved through a consensus-building process. The verification step helped to increase the reliability and validity of the analysis, as it ensured that the coding process was consistent and reliable across all coders. This qualitative analysis derived five main categories: \textit{Documentation-Related, Problems with Dependencies, Debugging, Conceptual Understanding, and
Implementation Issues}. These were further divided into subcategories as shown in Figure~\ref{fig:classification_questions}.

\begin{figure}[ht]
    \centering
    \includegraphics[width=3.5in]{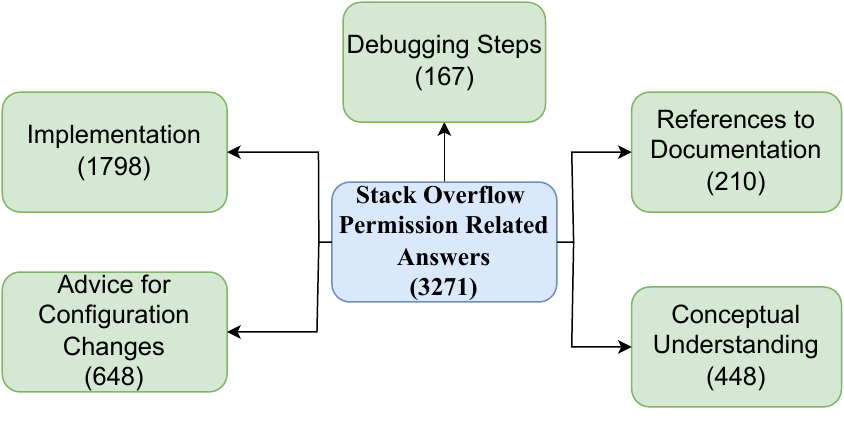}
    \caption{Categorization of Answers Corresponding to Provided Solutions }
    \label{fig:classification_answers}
\end{figure}

To address RQ2, similar to the analysis of RQ1, we used open coding and grouped the 3,271 answers into five major categories: \textit{Conceptual Understating,  References to Documentation, Debugging Steps, Implementation, and Advice for Configuration Changes}, as summarized in  Figure~\ref{fig:classification_answers}. 
Then, we mapped the question categories to answer categories, as summarized in Table ~\ref{table:mapping}.

Next, we analyzed the Stack Overflow users' characteristics for questions and answers. For each of the posted questions, community members can offer their solutions or leave them without a response. The proposed solutions are not always acceptable; thus community can express their opinion by voting up\footnote{\url{https://stackoverflow.com/help/privileges/vote-up}} (indicating approval) or down\footnote{\url{https://stackoverflow.com/help/privileges/vote-down}} (disagreeing with the correctness or usefulness of the solution). Among the proposed solutions, the author of the question eventually chooses the response that worked for them. This solution is labeled as the accepted answer\footnote{ \url{https://stackoverflow.com/help/accepted-answer}} to their question. We found 1,008 questions having accepted answers. Therefore, we got 1008 pairs of questions and accepted answers. For each of the proposed solutions, we analyze the profile of the Stack Overflow users, e.g., reputation and Stack Overflow
profile age (Table ~\ref{table:Developers_characteristics}). 
User reputation in Stack Overflow is a way to quantify the level of confidence other users have in one user's expertise, achieved by demonstrating that user's knowledge in a convincing manner\footnote{\url{https://stackoverflow.com/help/whats-reputation}}. The profile age of a Stack Overflow user indicates how long a user has been active on Stack Overflow. We calculated the average reputation score and profile age of the users for questions and answers (upvoted, downvoted, and accepted answers). This information will give us insight into the experience level of users who are asking questions and providing answers. The results are summarized in Table~\ref{table:Developers_characteristics}.

To address RQ3, we undertook a three-step process: firstly, we identified all the unique permissions that were referenced in the questions and answers of our dataset. After this identification process, we proceeded to categorize these permissions according to their protection level and their respective restriction type for Android 9/10, 11, 12, and 13 using the data collected in step \textcircled{3}.

In the second part of our analysis, we examined how these permissions evolved over time, effectively determining if their corresponding restriction type has changed or remained the same across different Android versions. The summary of questions and answers that discussed evolving permissions and those that dealt with non-evolving permissions are given in Table~\ref{table:permissionevolution}.

During our analysis, we observed that some questions and answers discussed several permissions, sometimes associated with different restriction types or protection levels. Thus, in the third part of the analysis, we counted the total number of permissions and their restriction types and protection levels in the questions and answers. We then calculated the number of questions and answers in each category.
In this classification, we used the associated Android version as indicated in a question or answer and the corresponding restriction lists. In this process, we encountered instances where the referenced permission for a particular Android version was not present in the official restriction list of that version. In such cases, we categorized these permissions as \textit{Mismatched}. For the protection level, we only analyzed SDK permissions. The analysis results are summarized in Table~\ref{table:permissionrestriction} and \ref{table:permissionprotection}.

To address RQ4, we explored how the information provided by developers in the questions, and answers aligns with the official Android documentation. If the protection level and restriction type were present, we compared whether the protection level and restriction type of permissions mentioned in the questions and answers were correct according to the official documentation. 

We also examined how the types of permissions mentioned in the questions align with those in the accepted answers. In some instances, we discovered that the accepted answer may refer to specific permissions, while in others, it might not. Specifically, we focused on understanding the restriction type of permissions in cases where both the question and the accepted answer discussed certain permissions.

\section{Results} 

\subsection{ \textbf{Results of RQ1:} Developers' Challenges related to Android Permissions  }

To aid our analysis, we derived a fine-grained categorization that groups developer's questions into the five broad categories (in green) shown in Figure~\ref{fig:classification_questions}. We further divided into 12 subcategories (in orange), nine sub-subcategories (in grey), and nine sub-sub-subcategories (in grey) shown in Figure~\ref{fig:classification_questions}. We complement each category with the corresponding example shown in Table~\ref{table:example}.



\subsubsection{Documentation-Related} 

The Android SDK documentation is an extensive and one-stop resource for official information on the Android Operating System. Developers' questions inquiring about documentation are categorized into \textit{Documentation-Related} category. We had 60 questions in this category. We further investigated them to understand what type of documentation-related challenges developers are facing. That insight is vital for any future revision and improvement in the official documentation. Based on that, we further divided them into three subcategories:

\textbf{\textit{Lack of Documentation:}} This category includes all questions indicating the lack of documentation necessary for proper usage of Android permissions. Interestingly, this is the largest subcategory with 35 questions indicating the difficulty developers face finding necessary documentation. For example, developer mention in Stack Overflow post that ``Android 13 - READ EXTERNAL STORAGE PERMISSION still usable (...)I didn’t find any information about this in the official documentation, but to be honest the documentation is focusing on media files only without any word about other file types(...).'' Therefore, there is a lack of documentation on using EXTERNAL STORAGE PERMISSION for the file types other than media files.

\textbf{\textit{Confusing Documentation:}} In this category, we include questions where the developers find the official documentation challenging to understand, unclear, and incomplete. The difference between \textit{Lack of Documentation} and \textit{Confusing Documentation} is that when there is no available official documentation related to specific permission and usage of permission on the other hand, \textit{Confusing Documentation} relates to any confusion or challenges related to existing documentation. There are 17 questions in this category, with most of them (70.0\%) left without an answer. Among those that provide responses are developers with higher reputations and experience.~\cite{mogavi2019hrcr} mentioned that a reputation score higher than 1,580.0 is a high reputation score on Stack Overflow. Table ~\ref{table:Developers_characteristics} shows that the average reputation score for users providing answers is 42,306.0, which falls into the high reputation category. On the other hand, the average reputation score for users asking questions is 1,844.7, which is also in the high reputation category but is close to the threshold of 1,580.0. This indicates that the users asking questions have less experience compared to the users providing answers.

We hypothesize that inexperienced Android developers find the permission-related documentation confusing, refraining from providing support to other developers.

\textbf{\textit{Contradictory Documentation:}} The subcategory includes questions that arise when different sources within the Android platform's documentation present contradictory information regarding API or permission actions. This subcategory had only eight questions, and three had an answer. 

For example,  DOWNLOAD\_WITHOUT\_NOTIFICATION permission is required for using ``VISIBILITY\_HIDDEN" value of the NotificationManager; however, the referenced permission is not listed in the AndroidManifest package index\footnotemark[7], suggesting it may not be available for third-party apps. 
This permission is similarly not present in the restriction lists. We were able to locate it though in the source code (AndroidManifest.xml\footnote{\url{https://android.googlesource.com/platform/packages/providers/DownloadProvider.git/+/master/AndroidManifest.xml}}) under NETWORK permission group. Note that  Google's official documentation for Manifest.permission NETWORK group\footnote{\url{https://developer.android.com/reference/android/Manifest.permission_group}} does not list the DOWNLOAD\_WITHOUT\_NOTIFICATION permission either.


Our findings show that \textit{Documentation-Related} challenges primarily stem from a \textit{Lack of Documentation} (35 instances). Although fewer challenges relate to \textit{confusing (17) and contradictory (8) Documentation}, they could lead to permission misuse. Thus, improved, detailed permission-related documentation is essential for enhancing app security.

\subsubsection{Problems with Dependencies} 
Within this category, we cover questions that revolve around the challenges faced by developers regarding permissions on particular devices, versions, or platforms. These difficulties often stem from the evolution of the Android system, resulting in diverse requirements and configurations that impact compatibility across different versions. In total, there are 246 questions in this category, with only 28.6\% of them having accepted answers. The challenges encountered by developers during this evolution serve as an indirect indication of inadequate documentation, i.e., existing documentation fails to offer sufficient support in handling compatibility across different versions and devices.

We further divided this category into three subcategories:  

\textbf{\textit{Version Dependency:}} This subcategory incorporates questions related to permissions that are specific to a particular version of Android OS. Often, the developers face these challenges due to evolving Android permission system that has been changing over time.  
Within this subcategory, consisting of 117 questions, developers consistently express a common challenge -  their app functions flawlessly across various versions except for one specific version.

\textbf{\textit{Device Dependency:}} This subcategory includes questions expressing difficulties with permissions on a particular Android device. 
These difficulties often arise as a result of the customizations or alterations that Original Equipment Manufacturers (OEMs) make to the Android operating system or device firmware. These modifications can include changes to the user interface, pre-installed apps, system settings, and security features. They can lead to challenges or issues related to permissions for developers working on a specific Android device. 
For example, the same set of permissions on Xiaomi devices can be treated differently than on Google mobile devices (Table ~\ref{table:example}). 
The issues related to incompatibility were previously raised by other studies. For example, ~\cite{silva2022saintdroid} and ~\cite{vieira2019gaindroid} discussed potential conflicts stemming from  API methods not being equally supported by different devices. In this category, we further explore these issues. This is a sizable category with 99 questions.

\textbf{\textit{Device and Version Dependency:}} Several developers experienced difficulties with specific Android versions on certain devices. To streamline the analysis process, we opted to examine these 30 questions separately, treating them as a distinct subcategory.


Among all \textit{Problems with Dependencies}-related challenges, issues tied to specific \textit{Versions} and \textit{Devices} are the most common. Addressing these challenges could involve providing version-specific and vendor-specific documentation on permissions and support tools to handle requirement and compatibility issues, thereby reducing \textit{Problems with Dependencies}.

\subsubsection{Debugging} 
In this category, we grouped questions related to unexpected behavior
and various errors in Android applications. This appeared to be the largest
category among the questions we collected, accounting for 54.0\%  (1,818)
of them. Among these 1,818 questions, 69.1\% have answers, and only 562 (31.0\%) have accepted answers. These questions were divided into two subcategories:

\textbf{\textit{Error Handling:}} 
This particular subcategory encompasses situations where developers aim to rectify an implementation that generates clear error messages. Within this subcategory, we have identified 1,419 questions where developers frequently encounter errors with specific error messages but struggle to understand the underlying cause to resolve the issue. The majority of these errors are directly related to incorrect permission usage, while some appear to stem from functionalities that involve permissions.
To better understand these issues, we have subdivided \textit{Debugging} into three categories:

  \begin{itemize}
   \item \textbf{\textit{Usage of Data Sensitive Permissions:}} Due to the sensitive nature of phone data, Google imposes restrictions on the access to sensitive data or information of the user\footnote{\url{https://support.google.com/googleplay/android-developer/answer/9888170}}. Google enforces these restrictions through various means, such as limiting certain permissions (e.g., dangerous and signature level permissions), requiring explicit approval for the usage of certain permissions (e.g., QUERY\_ALL\_PACKAGES permission), or mandating the implementation of a sensitive data policy.

According to Google's guidelines, these sensitive permissions should only be utilized when they are essential for the core functionality of the app. This should be explicitly stated in the app's data policy, which is presented to the user.


We classified 149 questions in this category and further divided it into two groups:

     \begin{itemize}
        \item \textbf{\textit{User Data Policy Declaration:}} Any inquiries or concerns pertaining to errors associated with the policy declaration of the sensitive permissions fall within this specific subcategory. 71.8\% of the questions related to \textit{Usage of Data Sensitive Permissions} fall into this subcategory.
        
        \item \textbf{\textit{Usage of Permissions without Core Functionality:}} 
        We group questions related to data-sensitive permissions that are declared but appear to be unnecessary for the core functionality of an app. For example, Google provides a warning indicating that used permissions are not necessary for even 1.0\% of the core functionality of the app (Table ~\ref{table:example}). There are 42 questions related to this subcategory.
    \end{itemize}

In the case of \textit{Usage of Data Sensitive Permissions}-related challenges, most are tied to \textit{User Data Policy Declaration} (71.8\%). This suggests developers often lack sufficient information, underestimate the importance of, or do not know how to declare a privacy policy, leading to their app's rejection by Google.
    
    \item \textbf{\textit{Absence of Required Permission:}} Within this subcategory, we have classified questions related to errors caused by missing permissions. 44.25\% of \textit{Error Handling}-related questions fall into this category. These inquiries typically involve a security exception error message generated by the system, indicating the absence of specific permissions necessary for the desired functionality.
    
    \item \textbf{\textit{Exception or Warning:}} While using permissions, developers commonly encounter problems that raise various exceptions, warnings, or error messages, although they do not provide an indication of how the error is associated with permissions.
    We group all these concerns in this generic category. 45.24\% of \textit{Error Handling}-related questions fall into this category.

    \end{itemize}

\textbf{\textit{Functional Improvement:}} This is the second category of questions related to debugging problems. In addition to encountering errors with the use of permissions, developers often seek to identify and resolve issues with a flawed implementation that lacks clear error messages. In such cases, developers seek to understand and rectify the issue. A total of 399 questions fall into this subcategory.

For the 1,818 questions in the \textit{Debugging} category, 78.1\% have specific error messages (1,419 questions from \textit{Error Handling}), and the rest of them are without (399 from \textit{Functional Improvement}). 
Out of 1,419 \textit{Error Handling} related questions, 777 errors (628 from \textit{Absence of Required Permission} and 149 from \textit{Usage of Data Sensitive Permissions}) are directly due to incorrect use or lack of permissions. If developers had sufficient documentation, guidelines, and understanding of the permission system, potentially over half (54.7\%) of these errors could be addressed. Handling errors not directly tied to permissions but arising from functionalities using them could also depend on the developer's experience, project knowledge, programming skills, as well as their understanding of documentation and the permission system.

\subsubsection{Conceptual Understating} This category comprises inquiries from developers seeking to understand concepts and justifications related to Android permissions (e.g., API usage, proposed changes in permission behavior, evolution in permission categorization). We further divided these inquiries into two subcategories: 

\textbf{\textit{Rationale Analysis:}} Occasionally, developers require an understanding of the underlying reasons for changes in permission behavior. The fact that there are only 57 questions related to this particular subcategory indicates that only a small number of developers demonstrate an interest in understanding the reasons behind these changes. 

 \textbf{\textit{Understanding
Permission Usage:}} Developers often seek to understand how, when, and where to use permissions. Among the 263 \textit{Conceptual Understanding} queries, 206 are specifically classified within this subcategory. This observation suggests that many developers may have limited knowledge of the permission system, prompting them to inquire about the application and management of permissions.

\subsubsection{Implementation Issues} 
With 28.2\% (940) of all the questions, this category stands as the second largest. The questions in this category focus on various aspects of problems related to the implementation of specific functionality or features. These questions were further divided into two subcategories: 

\textbf{\textit{Functionality Implementation:}} In this subcategory, we have classified the questions that focus on developers seeking guidance on the proper implementation of specific functionalities. A total of 683 questions have been categorized in this subcategory. We have identified that some of these questions are specifically directed toward features, while others seek implementation guidelines for using specific permissions. We further divided these questions into two subcategories accordingly:
\begin{itemize}

\item\textbf{\textit{Feature Specific:}} 182 questions related to specific types of feature implementations are categorized in this category. Based on the permission usage in specific feature implementation, we further divided the questions into the following subcategories: 
\begin{itemize}

    \item \textbf{\textit{Location Services:}} The questions related to the functionality in Android apps that requires or uses geographical location data. It involves the management of permissions necessary for accessing GPS and the device's location (such as approximate location, precise location, or both). There are 19 questions related to location services.
    
   \item \textbf{\textit{Data \& File Management: }}The questions discuss handling, storing, and retrieving data on Android devices while adhering to required permissions. This includes managing data files, databases, and shared preferences, with challenges often arising from permission requests, privacy constraints, and data security. There are 60 questions in this category.  

    \item \textbf{\textit{Services \& Background Processes:}} A `service' is a component that runs in the background to perform long-running tasks. The questions in this category aim to obtain guidance regarding the utilization of permissions associated with these long-running tasks, such as file downloads or music playback, that extend beyond the user interface. It is crucial to ensure the secure usage of these services since they often remain hidden from users. 
    There are 22 questions grouped in this category. 
    
    \item \textbf{\textit{Bluetooth \& Wifi Connectivity in Android: }} Within this category, we combined questions that revolve around the utilization and administration of wireless connections within Android apps, with a specific focus on Bluetooth and Wifi. This encompasses the creation, management, and implementation of features related to Bluetooth or Wifi connections, handling connectivity permissions, and ensuring secure data transfer and device pairing. This group includes 19 questions.  

    \item \textbf{\textit{UI \& Navigation:}} The questions included in this subcategory encompass topics concerning the User Interface (UI) and navigation features in Android apps. They may involve permissions for overlaying windows, modifying the corresponding system settings (e.g., screen-recording or screen-sharing, chat head, etc.), or accessing services that affect the UI.

    \item \textbf{ \textit{Camera \& Media Access on Android: }}The questions inquiring about the app access to the device's camera and media files, such as photos and videos, are stored in the device's storage or SD card. Only 23 questions refer to the camera or media-related features. 
    
    \item \textbf{\textit{Device \& Hardware Interaction:}} The questions in this subcategory revolve around the use of necessary permissions for access to Android hardware components (e.g., sensors).
  There are 16 questions related to this category. This is the category with the least number of questions.
    \end{itemize}

In the \textit{Feature Specific}, developers face most challenges related to \textit{Data \& File Management}. This is one of the crucial feature which can be exploited if permissions are not used correctly for this feature. 

\item \textbf{\textit{Permissions management:}}
This category includes questions that delve into the strategies and techniques employed for requesting and managing permissions. Since we treat it as a generic category,  it is important to note that these questions do not specifically mention any particular type of feature.


\end{itemize}

\textbf{\textit{Alternative Solutions:}}  As opposed to the functionality implementation category,  the questions in this category address aspects related to achieving more efficient and/or secure solutions. Developers seek alternatives to existing solutions and may specifically request guidance on optimal approaches. 256 questions were related to this subcategory. This subcategory was divided into five subcategories based on the specific reasons for seeking alternative solutions:

\begin{itemize}
 
    \item \textbf{\textit{Adapting Permission Policy \& Changes: }}The questions in this category revolve around situations where developers seek to modify their code in response to changes in the Android permission system. These questions often refer to soon-to-be deprecated permissions and their alternatives.

    This subcategory has the highest (158) number of questions among 256 \textit{Adapting Permission Policy and Changes}-related questions.

    \item \textbf{\textit{Improved Solutions:}} This category includes 57 questions from developers seeking better solutions. Generally, these developers are already familiar with at least one existing solution but are seeking ways to enhance it.

    \item \textbf{\textit{Avoiding Permission Usage: }}  In this category, we grouped questions inquiring about an alternative solution that does not involve utilizing permissions. Often times developers do not specify a particular reason for their requests; thus, the underlying motive for avoiding permission usage remains unclear. It may stem from various factors, such as simplifying the development process, improving the user experience by minimizing unnecessary permission requests, and so on.
    We identified 36 questions that fall in this category.  
    

    \item \textbf{\textit{Power Consumption:}} In this category, we have questions that revolve around developers seeking to modify their code to minimize power consumption (e.g., optimize the battery consumption). With only five questions, this is the sub-subcategory with the lowest number of questions.

    

    From the above sub-subcategories of \textit{Alternative Solutions}, we found that 61.7\% (158 questions out of 256) questions seek alternative solutions for adapting the changes in the Android permission system. Therefore, developers often feel challenging to cope with the change and updates in the Android permission system.    
    
\end{itemize}

The results above show that, out of all the challenges, 28.2\% (or 940 questions) fall under \textit{Implementation Issues} category. Most often, these challenges are connected to \textit{Functionality Implementation.} This mainly includes strategies and techniques for requesting and handling permissions and for managing access control in an Android app, which we refer to as \textit{Permission \& Access Control.} In certain instances, developers seek \textit{Alternative Solutions,} typically as a response to changes in the permission system. Out of the 940 questions in the \textit{Implementation Issues} category, 64.0\% have received answers. Only 27.3\% of those \textit{Implementation Issues} have been marked as accepted answers, which is the lowest among all the categories. This suggests that developers often struggle to find satisfactory solutions from the community for issues related to \textit{Implementation Issues.}

\begin{tcolorbox}
\textbf{Summary of RQ1:} We have classified the challenges faced by developers into five main categories, further divided into 12 subcategories, nine sub-subcategories, and nine sub-sub-subcategories. We found that the most common type of challenge faced by developers involves \textit{Debugging.} In fact, 78.1\% of these challenges include a specific error message. It's interesting to note that about 54.7\% of these errors are directly linked to the misuse or incorrect app of Android permissions. The second most common type of challenge is related to \textit{Implementation Issues.} In this category, developers primarily request assistance with \textit{Functionality Implementation} and seek \textit{Alternative Solutions.} Notably, 61.7\% of requests for \textit{Alternative Solutions} arise due to changes in the Android permission policy, highlighting that developers often find it hard to keep up with the evolving permission systems. Even though questions related to the \textit{Documentation-Related} and \textit{Problems with Dependencies} categories are less frequent, reducing these challenges is critically important in reducing misuse of permissions and avoiding compatibility issues.

Finally, out of the 3,327 questions, only 30.2\% have accepted answers. This is significantly lower than the overall acceptance rate of 53.0\% on Stack Overflow, as referenced in a study by ~\cite{yazdaninia2021characterization}. This indicates that developers are not receiving adequate assistance from the community in addressing their challenges related to permissions.
\end{tcolorbox}

\subsection{ \textbf{Results of RQ2:} Solutions Provided to Android Permission-Related Challenges} To aid our analysis, we categorized 3,271 corresponding responses to the  3,327 questions in five main categories, which are summarized in Figure~\ref{fig:classification_answers}, and mapped to the questions' categories in Table ~\ref{table:mapping}. We also complement each of these categories with the corresponding example
shown in Table ~\ref{table:example_solution}.

\textbf{\textit{Conceptual Understanding:}} The answers explain the inner workings of principles and concepts of using permission. For example, the answer provided the understanding of when the feature works without permission and with permission (Table ~\ref{table:example_solution}). There are 484 solutions providing \textit{Conceptual Understating}. 

\textbf{\textit{References to Documentation:}} All community responses that provide references to official Android documentation or learning resources to resolve issues are included in this category. Only 210 solutions are related to this category. 

\textbf{\textit{Debugging Steps:}} As opposed to implementation guidance,  community responses in this category explain the necessary steps to debug problematic code or situation to further understand the root cause of the issue. Even though there are a lot of challenges related to \textit{Debugging} (1,818), there are only 167 responses that give steps on how to debug.

\textbf{\textit{Implementation:}} The solutions provide advice for implementing a functionality, and its improvement or solutions to errors, warnings, and exceptions fall into this category. These solutions typically include code examples. Out of a total of 3,271 solutions, 1,798 of them (54.9\%) fall under the \textit{Implementation} category. This indicates that the developer community often provides practical \textit{Implementation} instructions. This is the case whether the issue at hand relates to implementation, debugging, conceptual understanding, or dependencies. Instead of focusing on identifying the root cause of an error or warning, the Stack Overflow community tends to provide implementations as solutions directly. This means they often offer steps to solve a problem rather than exploring why the problem occurred in the first place.

\textbf{\textit{Advice for Configuration Changes:}} In this category, we include answers that provide advice for changing configuration settings for various API levels to obtain desired behavior or functionality. This category has 648 answers, making it the second most common type of solution. This means that the Stack Overflow community often provides advice on how to adjust to the different requirements and configurations of various Android versions and devices.

\begin{table}[ht]
\centering
\scalebox{0.68}{
\begin{tabular}{lrrrrr}
\hline
\multirow{2}{*}{\textbf{\begin{tabular}[c]{@{}l@{}}Questions' \\ Categories\end{tabular}}} & \multicolumn{5}{c}{\textbf{\#Answers}} \\ \cline{2-6} 
 &
  \multicolumn{1}{c}{\begin{tabular}[c]{@{}c@{}}Implementation\end{tabular}} &
  \multicolumn{1}{c}{\begin{tabular}[c]{@{}c@{}}Advice for \\ Configuration\\Changes\end{tabular}} &
  \multicolumn{1}{c}{\begin{tabular}[c]{@{}c@{}}References to \\ Documentation\end{tabular}} &
  \multicolumn{1}{c}{\begin{tabular}[c]{@{}c@{}}Conceptual \\ Understanding\end{tabular}} &
  \multicolumn{1}{c}{\begin{tabular}[c]{@{}c@{}}Debugging \\ Steps\end{tabular}} \\ \hline
  \begin{tabular}[l]{@{}c@{}}
Documentation -Related \end{tabular}                                                                     & 21    & 10  & 11  & 15 & -  \\
  \begin{tabular}[c]{@{}l@{}}Conceptual Understanding\end{tabular}               & 71    & 37  & 32  & 107   & 3 \\
  \begin{tabular}[l]{@{}c@{}} Problems with Dependencies \end{tabular}                                                                        & 110   & 62  & 10  & 18  & 12  \\
  \begin{tabular}[l]{@{}c@{}}
Debugging   \end{tabular}                                                                       & 1,050  & 383 & 99  & 170 & 150 \\
\begin{tabular}[l]{@{}c@{}}Implementation Issues   \end{tabular}                                                                & 546   & 156 & 58  & 138 & 2   \\ \hline
Total                                                                             & 1,798  & 648 & 210 & 448 & 167 \\ \hline
\end{tabular}}
\caption{\label{table:mapping} Mapping of Questions' and Answers' Categories}
\end{table}

\begin{table}[ht]
\centering
\scalebox{0.8}{
\centering
\begin{tabular}{lrrrrr}
\hline
\multicolumn{1}{c}{\textbf{\begin{tabular}[c]{@{}c@{}}Users'\\ Profile\end{tabular}}} &
  \multicolumn{1}{c}{\textbf{Questions}} &
  \multicolumn{1}{c}{\textbf{Answers}} &
  \multicolumn{1}{c}{\textbf{\begin{tabular}[c]{@{}c@{}}Accepted   \\ Answers\end{tabular}}} &
    \multicolumn{1}{c}{\textbf{\begin{tabular}[c]{@{}c@{}}Up Voted   \\ Answers\end{tabular}}} &
  \multicolumn{1}{c}{\textbf{\begin{tabular}[c]{@{}c@{}}Down Voted   \\ Answers\end{tabular}}} \\ \hline
Average Reputation &
  1,844.7 &
  42,306.1 &
  62,316.0 &
  46,868.4 &
  14,302.0 \\
Average Profile Age &
  6.3 years &
  7.2 years &
  7.4 years &
  7.3 years &
  5.2 years \\ \hline
\end{tabular}}
\caption{ User Profiles for the Users Asking Permission-Related Questions, Providing Accepted, Up Voted, and Down Voted Answers
\label{table:Developers_characteristics}}
\end{table}





We mapped the developers' questions to the corresponding response categories. The results are shown in Table~\ref{table:mapping}.
It is clear that for all types of developers' questions, the most common type of solution provided by the community is related to \textit{Implementation}. Interestingly, despite \textit{Debugging} being the most common problem developers face, solutions offering \textit{Debugging Steps} are the least common. This shows that the solutions provided often do not align with the problems faced by developers. This mismatch could be the reason why there is such a low number of accepted answers.

We further analyzed users' characteristics to understand which type of users are providing solutions to the permission-related challenges. Based on the information in Table ~\ref{table:Developers_characteristics}, users who provide accepted and up voted answers on Stack Overflow generally have a higher reputation score, with an average of 62,316.0 and 46,868.4 respectively. This means more experienced users on Stack Overflow are providing more useful and correct answers. On the other hand, the average reputation score for users who ask questions is significantly lower, coming last at an average of 1,844.7 scores, and the users having the second lowest reputation score (an average of 14,302.0 scores) are providing down voted answers. 

This data suggests that answers accepted and valued by the community often come from more experienced users. Conversely, less experienced users, indicated by their lower reputation scores, usually ask questions about permission challenges and offer less helpful solutions. This trend suggests that less experienced developers may face more challenges related to permissions. These challenges could stem from their lack of understanding or knowledge about the permission system, which might include things like levels of permission protection, restriction types, and more. One possible cause for these issues could be the absence of clear and comprehensive documentation that these developers can use to gain a better understanding of the system.

\begin{table}[ht]
\centering
\begin{footnotesize}
\begin{tabular}{|l|l|l|} 
\hline
Category & Example \\ 
\hline
\begin{tabular}[c]{@{}l@{}}Implementation\end{tabular} & \begin{tabular}[c]{@{}l@{}}You can handle it in WebViewClient like codes below, \\but it will replace the whole page of WebView with \\your error page: (...), Alternatively, you can handle \\the error in Html, you can define an error window \\with the same size of your webview, and hide/show \\it according to the internet states:(...), Notes: you will \\have to define following permission to have\\navigator.onLine work:(...)\end{tabular}\\ 
\hline
\begin{tabular}[c]{@{}l@{}}Advice for \\Configuration Changes\end{tabular} & \begin{tabular}[c]{@{}l@{}}Change this:compile \`{}com.google.firebase:firebase-\\core:12.0.0' to this: compile \`{}com.google.firebase:\\firebase-core:12.0.1'(...)\end{tabular} \\ 
\hline
 \begin{tabular}[c]{@{}l@{}}References to \\Documentation\end{tabular} & \begin{tabular}[c]{@{}l@{}}Use the FingerprintManagerCompat instead, that \\was handling permissions correctly for me.See: \\https://developer.android.com/reference/android(...)\end{tabular} \\ 
\hline
\begin{tabular}[c]{@{}l@{}}Conceptual \\Understanding\end{tabular} & \begin{tabular}[c]{@{}l@{}}I have had instances where a user had a third-party \\file manager installed (File Manager+) and in those \\cases reading from the Uri returned by ACTION\_GET\\
\_CONTENT would fail with a permission \\error if the READ\_EXTERNAL\_STORAGE permission \\was not first granted (only if they used the third-party \\app to select the file, if they used Google Drive or the \\normal system selection it worked fine without the \\permission).\end{tabular} \\ 
\hline
Debugging Steps & \begin{tabular}[c]{@{}l@{}}InvocationTargetException is thrown because Reflection \\wraps any other Exception in the underlying method into \\this class. Try putting the whole code in a try/catch block \\and printing the stack trace with e.printStackTrace(). \\There should be a "Caused By:" line which will hopefuly \\point you to the right direction.(...)\end{tabular} \\
\hline
\end{tabular}

\caption{\label{table:example_solution} Examples of Responses Offered by the Community for Each Category}
\end{footnotesize}
\end{table}

\begin{tcolorbox}
\textit{\textbf{Summary of RQ2: }We have classified the challenges faced
by developers into five main categories. Interestingly, despite \textit{Debugging} being the most common problem developers face, solutions offering \textit{Debugging Steps} are the least common. On the other hand, the majority of answers,
54.9\%, provide \textit{Implementation}-related solutions. This suggests a disconnect between the issues developers encounter and the solutions offered by the community, which could explain why there is such a low percentage (30.2\%) of accepted answers. The Stack Overflow community often provides advice on implementing solutions rather than identifying the root cause of a problem.}
\end{tcolorbox}

\subsection{\textbf{Results of RQ3:} Specific Restriction Types and Protection Levels of Permissions that Tend to Cause the most Issues }

The results of our analysis are presented in Tables~\ref{table:permissionevolution}, ~\ref{table:permissionrestriction},and~\ref{table:permissionprotection}.  

Out of  765 existing permissions, developers sought guidance for 185  permissions. Table~\ref{table:permissionrestriction} presents the discussed permissions with regard to their restriction types. In our analysis, we came across questions and answers that discussed multiple permissions, often with different restriction types and protection levels. To account for these situations,  these questions and answers were considered in each mentioned restriction type or protection level category. 

While the majority of questions dealt with the SDK permissions, a small portion of questions (9) referred to conditionally blocked permissions, and one question to blocked permission. Both types of permissions are restricted, conditionally blocked to the app’s target API level, and blocklisted permissions cannot be used regardless of the app’s target API level\footnotemark[5]. 
We further investigated 21 unique conditionally blocked permissions and found that 18 of those permissions were usable till Android 8 (Conditionally blocked-max-o), and three of those permissions were usable till Android 11 (Conditionally blocked-max-r). The related questions did not mention any Android version. Therefore, it was not possible to verify whether these permissions were usable for the specific cases referred to in those questions.

To our surprise, the vast majority of questions (81.0\%) were regarding non-evolving permissions, which means these permissions remained stable across four recent Android versions (Table~\ref{table:permissionevolution}). Further analysis revealed that 157 of these permissions belonged to the SDK group, indicating that they are officially documented and supported for use by third-party developers. This finding suggests that the official documentation may not provide sufficient information for developers, prompting them to seek additional guidance.

 In comparison, only 13 questions included inquiries about 11 evolving permissions, and only nine answers discussed ten evolving permissions. The details of these evolving permissions are summarized in Tables~\ref{table:evolving_question} and~\ref{table:evolving_answer}. Since these questions and answers did not indicate Android versions, it is difficult to pinpoint the exact type of restrictions that developers find challenging. However, we observe a clear pattern, i.e., permissions from restricted categories over time are moved to less restricted categories. Most of these cases involve unsupported and conditionally blocked permissions. Hence, developers struggle with permissions whose availability and treatment depend on specific Android versions.

\begin{table}[ht]
\centering
\scalebox{0.78}{
\begin{tabular}{lrrrr}
\hline
\multirow{2}{*}{\textbf{\begin{tabular}[c]{@{}l@{}}Evolution\\ Criteria\end{tabular}}} &
  \multicolumn{2}{c|}{\textbf{Questions}} &
  \multicolumn{2}{c}{\textbf{Answers}} \\
 &
  \multicolumn{1}{c}{\#Unique Permissions} &
  \multicolumn{1}{c|}{\#Questions} &
  \multicolumn{1}{c}{\#Unique Permissions} &
  \multicolumn{1}{c}{\#Answers} \\ \hline
Non-Evolving      & 175 (94.5\%) & 2,696 (81.0\%) & 123 (93.1\%) & 1,230 (37.6\%) \\
Evolving  & 10 (5.4\%)  & 13 (0.3\%)   & 10 (7.5\%)  & 9 (0.2\%)  \\
Both  & -  & 16 (0.4\%)  & -   & 9 (0.2\%)   \\
No Permission & -   & 603 (18.1\%)  & -   & 2023 (61.8\%) \\ \hline
Total         & 185 & 3,327 & 132 & 3,271 \\ \hline
\end{tabular}}
\caption{The Summary of the Collected Set }
\label{table:permissionevolution} 
\end{table}


\begin{table}[ht]
\centering
\scalebox{0.68}{
\begin{tabular}{lrrrrrr}
\hline
\multirow{2}{*}{\textbf{Restriction Type}} & \multicolumn{3}{c|}{\textbf{Questions}} & \multicolumn{3}{c}{\textbf{Answers}} \\ \cline{2-7} 
 & \multicolumn{1}{c}{\#Permissions} & \multicolumn{1}{c}{\begin{tabular}[c]{@{}c@{}}\#Unique\\ Permissions\end{tabular}} & \multicolumn{1}{c}{\#Questions} & \multicolumn{1}{l}{\#Permissions} & \multicolumn{1}{l}{\begin{tabular}[c]{@{}l@{}}\#Unique\\ Permissions\end{tabular}} & \multicolumn{1}{l}{\#Answers} \\ \hline
SDK & 5,537 & 157 & 2,717 & 2,033 & 106 & 1,216 \\
Conditionally blocked & 95 & 21 & 9 & 28 & 13 & 24 \\
Blocklist & 1 & 1 & 1 & 1 & 1 & 1 \\
Unsupported & 0 & 0 & 0 & 0 & 0 & 0 \\
Mismatched & 52 & 6 & 13 & 57 & 12 & 40 \\ \hline
Total & 5,685 & 185 & 2,740 & 2,119 & 132 & 1,281 \\ \hline
\end{tabular}}
\caption{The Distribution of Permissions Restrictions within the Collected Set }
\label{table:permissionrestriction} 
\end{table}

\begin{table}[ht]
\centering
\scalebox{0.68}{
\begin{tabular}{lrrrrrr}
\hline
\multicolumn{1}{c}{\multirow{2}{*}\textbf{{\begin{tabular}[c]{@{}c@{}}Protection\\ Level\end{tabular}}}} & \multicolumn{3}{c|}{\textbf{Questions}} & \multicolumn{3}{c}{\textbf{Answers}} \\ \cline{2-7} 
\multicolumn{1}{c}{} & \multicolumn{1}{c}{\#Permissions} & \multicolumn{1}{c}{\begin{tabular}[c]{@{}c@{}}\#Unique\\ Permissions\end{tabular}} & \multicolumn{1}{c}{\#Questions} & \multicolumn{1}{l}{\#Permissions} & \multicolumn{1}{l}{\begin{tabular}[c]{@{}l@{}}\#Unique\\ Permissions\end{tabular}} & \multicolumn{1}{l}{\#Answers} \\ \hline
Signature & 418 & 18 & 354 & 123 & 12 & 102 \\
Normal & 1,420 & 52 & 817 & 301 & 34 & 225 \\
Dangerous & 3,492 & 51 & 1,827 & 1,546 & 39 & 819 \\
\begin{tabular}[c]{@{}l@{}}SignatureOrSystem \end{tabular} & 207 & 36 & 181 & 63 & 21 & 61 \\\hline
Total & 5,537 & 157 & 3,179 & 2,033 & 106 & 1,207 \\ \hline
\end{tabular}}
\caption{Protection Level of the  SDKs Permission. }
\label{table:permissionprotection} 
\end{table}


\begin{table}[H]
\centering
\scalebox{0.64}{
\begin{tabular}{lllll}
\hline
\multirow{2}{*}{\textbf{Permission}} & \multicolumn{3}{c}{\textbf{Restriction Types}} \\ \cline{2-5} &  Android 9/10 & Android 11 & Android 12 & Android 13 \\
\hline

MANAGE\_DEVICE\_ADMINS & Unsupported & Unsupported & Unsupported & SDK \\

WRITE\_SMS & \begin{tabular} [c]{@{}l@{}} Conditionally \\Blocked-max-o\end{tabular} & \begin{tabular} [c]{@{}l@{}} Conditionally \\Blocked-max-o\end{tabular} &\begin{tabular} [c]{@{}l@{}} Conditionally \\Blocked-max-o\end{tabular} & SDK \\

READ\_FRAME\_BUFFER & Unsupported & Unsupported & \begin{tabular} [c]{@{}l@{}} Conditionally \\Blocked-max-r\end{tabular} & \begin{tabular} [c]{@{}l@{}} Conditionally \\Blocked-max-r\end{tabular} \\

START\_ACTIVITIES\_FROM\_BACKGROUND & Blocklist & Blocklist & SDK & SDK \\

REQUEST\_NETWORK\_SCORES & \begin{tabular} [c]{@{}l@{}} Conditionally \\Blocked-max-o\end{tabular} & SDK & SDK & SDK \\

NETWORK\_SETTINGS & \begin{tabular} [c]{@{}l@{}} Conditionally \\Blocked-max-o\end{tabular} & SDK & SDK & SDK \\

NETWORK\_STACK & \begin{tabular} [c]{@{}l@{}} Conditionally \\Blocked-max-o\end{tabular} & SDK & SDK & SDK \\

CAPTURE\_VIDEO\_OUTPUT & Unsupported & Unsupported & \begin{tabular} [c]{@{}l@{}} Conditionally \\Blocked-max-r\end{tabular} & \begin{tabular} [c]{@{}l@{}} Conditionally \\Blocked-max-r\end{tabular} \\

READ\_PRECISE\_PHONE\_STATE & \begin{tabular} [c]{@{}l@{}} Conditionally \\Blocked-max-o\end{tabular} & SDK & SDK & SDK \\

CAPTURE\_SECURE\_VIDEO\_OUTPUT & Unsupported & Unsupported & \begin{tabular} [c]{@{}l@{}} Conditionally \\Blocked-max-r\end{tabular} & \begin{tabular} [c]{@{}l@{}} Conditionally \\Blocked-max-r\end{tabular} \\
\hline
\end{tabular}}
\caption{Summary of Evolving Permissions in Questions}
\label{table:evolving_question}
\end{table}


\begin{table}[htbp]
\centering
\scalebox{0.64}{
\begin{tabular}{lllll}
\hline
\multirow{2}{*}{\textbf{Permission}} & \multicolumn{3}{c}{\textbf{Restriction Types}} \\ \cline{2-5} & Android 9/10 & Android 11 & Android 12 & Android 13 \\
\hline
WRITE\_SMS & \begin{tabular} [c]{@{}l@{}} Conditionally \\Blocked-max-o\end{tabular} & \begin{tabular} [c]{@{}l@{}} Conditionally \\Blocked-max-o\end{tabular} & \begin{tabular} [c]{@{}l@{}} Conditionally \\Blocked-max-o\end{tabular} & SDK \\
MANAGE\_DEVICE\_ADMINS & Unsupported & Unsupported & Unsupported & SDK \\
NETWORK\_SETTINGS & \begin{tabular} [c]{@{}l@{}} Conditionally \\Blocked-max-o\end{tabular} & SDK & SDK & SDK \\
READ\_FRAME\_BUFFER & Unsupported & Unsupported & \begin{tabular} [c]{@{}l@{}} Conditionally \\Blocked-max-r\end{tabular} & \begin{tabular} [c]{@{}l@{}} Conditionally \\Blocked-max-r\end{tabular} \\
CAPTURE\_VIDEO\_OUTPUT & Unsupported & Unsupported & \begin{tabular} [c]{@{}l@{}} Conditionally \\Blocked-max-r\end{tabular} & \begin{tabular} [c]{@{}l@{}} Conditionally \\Blocked-max-r\end{tabular} \\
CAPTURE\_SECURE\_VIDEO\_OUTPUT & Unsupported & Unsupported & \begin{tabular} [c]{@{}l@{}} Conditionally \\Blocked-max-r\end{tabular} & \begin{tabular} [c]{@{}l@{}} Conditionally \\Blocked-max-r\end{tabular} \\
START\_ACTIVITIES\_FROM\_BACKGROUND & Blocklist & Blocklist & SDK & SDK \\
NETWORK\_STACK & \begin{tabular} [c]{@{}l@{}} Conditionally \\Blocked-max-o\end{tabular} & SDK & SDK & SDK \\
REQUEST\_NETWORK\_SCORES & \begin{tabular} [c]{@{}l@{}} Conditionally \\Blocked-max-o\end{tabular} & SDK & SDK & SDK \\
\hline
\end{tabular}}
\caption{Summary of Evolving Permissions in Answers}
\label{table:evolving_answer}
\end{table}




Developers discussed 185 unique permissions a total of 5685 times in the questions. 
The most discussed permissions in the questions are summarized in Table ~\ref{table:most_permission_question}. We observed that most of the required permissions fall into the Dangerous category for both questions and answers. These permissions typically involve storage management, location tracking, or access to sensitive device resources and data (e.g., camera, reading the contact list, or phone's state).  

\begin{table}[htbp]
\centering
\scalebox{0.46}{
\begin{tabular}{lrlllrll}
\hline
\multicolumn{4}{c|}{\textbf{Questions}} &\multicolumn{4}{c}{\textbf{Answers}}\\\hline
Permission & Frequency & \begin{tabular}[c]{@{}l@{}}Protection\\ Level\end{tabular} & \begin{tabular}[c]{@{}l@{}}Protection\\ Level\end{tabular} & Permission & Frequency & \begin{tabular}[c]{@{}l@{}}Protection \\Level \end{tabular} & \begin{tabular}[c]{@{}l@{}}Protection\\ Level\end{tabular} \\
\hline
WRITE\_EXTERNAL\_STORAGE & 648 & Dangerous & SDK & WRITE\_EXTERNAL\_STORAGE & 246 & Dangerous & SDK \\
READ\_EXTERNAL\_STORAGE & 525 & Dangerous & SDK & ACCESS\_FINE\_LOCATION & 214 & Dangerous & SDK \\
ACCESS\_FINE\_LOCATION & 457 & Dangerous & SDK & READ\_EXTERNAL\_STORAGE & 170 & Dangerous & SDK \\
INTERNET & 395 & Normal & SDK & ACCESS\_COARSE\_LOCATION & 146 & Dangerous & SDK \\
CAMERA & 297 & Dangerous & SDK & CAMERA & 140 & Dangerous & SDK \\
ACCESS\_COARSE\_LOCATION & 284 & Dangerous & SDK & ACCESS\_BACKGROUND\_LOCATION & 79 & Dangerous & SDK \\
READ\_PHONE\_STATE & 185 & Dangerous & SDK & READ\_PHONE\_STATE & 69 & Dangerous & SDK \\
MANAGE\_EXTERNAL\_STORAGE & 183 & Signature & SDK & INTERNET & 68 & Normal & SDK \\
ACCESS\_NETWORK\_STATE & 180 & Normal & SDK & READ\_CONTACTS & 66 & Dangerous & SDK \\
BLUETOOTH & 142 & Normal & SDK & MANAGE\_EXTERNAL\_STORAGE & 6 & Signature & SDK \\
\hline
\end{tabular}}
\caption{Frequency of Top 10 Discussed Permissions}
\label{table:most_permission_question}
\end{table}


For the questions, we observed a small portion of questions (13) that inquired about permissions absent in the Android versions mentioned in these questions (Table~\ref{table:permissionrestriction}). These \textit{Mismatched} questions discussed 52 permissions (six unique permissions) summarized in Table~\ref{table:mismatched_question}. These permissions are classified as either Dangerous or Signature in terms of their protection level, indicating a higher risk they imply. As stated explicitly in the Android documentation, one of the permissions, MANAGE\_WIFI\_INTERFACES\footnote{\url{https://developer.android.com/reference/android/Manifest.permission\#MANAGE_WIFI_INTERFACES}},  cannot be used by third-party developers but did not mention any protection level. This inquiry indicates the lack of developers understanding of permission restrictions.  

\begin{table}[htbp]
\scalebox{0.65}{
\centering
\begin{tabular}{lrrrll}
\hline
\textbf{Permission Name} & \textbf{\begin{tabular}[c]{@{}l@{}}Mismatched \\Version\\Stated in \\Question\end{tabular}}  & \textbf{\begin{tabular}[c]{@{}l@{}}Available\\ in Version\end{tabular}} & \textbf{\#Permissions} &\textbf{\begin{tabular}[c]{@{}l@{}}Restriction \\Type\end{tabular}} & \textbf{\begin{tabular}[c]{@{}l@{}}Protection \\Level\end{tabular}} \\
\hline
BLUETOOTH\_SCAN & 11 & 12 & 12  & SDK & Dangerous \\
BLUETOOTH\_CONNECT & 11 & 12 & 15  & SDK & Dangerous \\
BLUETOOTH\_ADVERTISE & 11 & 12 & 2  & SDK & Dangerous \\
POST\_NOTIFICATIONS & 12 & 13 & 1  & SDK & Dangerous \\
MANAGE\_EXTERNAL\_STORAGE & 10 & 11 & 21  & SDK & Signature \\
MANAGE\_WIFI\_INTERFACES & 12 & 13 & 1  & SDK & - \\
\hline
\end{tabular}}
\caption{Summary of Mismatched Permissions in Questions}
\label{table:mismatched_question}
\end{table}


Similarly, Table~\ref{table:mismatched_answer} shows the summary of the similar analysis performed for answers. A small portion of answers (nine) discussed 57 (12 unique)
\textit{Mismatched} permissions that appeared to be absent in the Android version mentioned in these answers. 
Similarly to the mismatched permissions present in questions, the \textit{Mismatched} permissions in answers are also Dangerous or Signature. 

\begin{table}[htbp]
\scalebox{0.64}{
\centering
\begin{tabular}{lrrrll}
\hline
\textbf{Permission Name} & \textbf{\begin{tabular}[c]{@{}l@{}}Mismatched \\Version\\Stated in \\Question\end{tabular}}  & \textbf{\begin{tabular}[c]{@{}l@{}}Available\\ in Version\end{tabular}} & \textbf{\#Permissions} &\textbf{\begin{tabular}[c]{@{}l@{}}Restriction \\Type\end{tabular}} & \textbf{\begin{tabular}[c]{@{}l@{}}Protection \\Level\end{tabular}} \\
\hline
READ\_MEDIA\_VIDEO & 10 & 13 & 3  & SDK & Dangerous \\
READ\_MEDIA\_IMAGES & 10 & 13 & 3  & SDK & Dangerous \\
READ\_MEDIA\_AUDIO & 10 & 13 & 3  & SDK & Dangerous \\
MANAGE\_EXTERNAL\_STORAGE & 10 & 11 & 9  & SDK & Signature \\
BLUETOOTH\_SCAN & 11 & 12 & 10  & SDK & Dangerous \\
BLUETOOTH\_CONNECT & 11 & 12 & 10& SDK & Dangerous \\
QUERY\_ALL\_PACKAGES & 10 & 11 & 6  & SDK & Normal \\
POST\_NOTIFICATIONS & 12 & 13 & 3  & SDK & Dangerous \\
BLUETOOTH\_ADVERTISE & 11 & 12 & 3  & SDK & Dangerous \\
MANAGE\_MEDIA & 11 & 12 & 2  & SDK & Signature \\
MANAGE\_ONGOING\_CALL & 11 & 12 & 2  & SDK & Signature\\
HIGH\_SAMPLING\_RATE\_SENSORS & 11 & 12 & 2  & SDK & Normal \\
\hline
\end{tabular}}
\caption{Summary of Mismatched Permissions in Answers} 
\label{table:mismatched_answer}
\end{table}


The summary of discussed SDK permissions with regards to their protection level is presented in Table~\ref{table:permissionprotection}. The protection levels are only available for officially documented SDK permissions.
Similarly to questions, dangerous permissions were the most discussed in the provided answers (819). 

Among these 1,827 questions discussing dangerous permissions, 1,140 are related to \textit{Debugging}-related challenges followed by \textit{Implementation Issues}(415). This is not surprising as dangerous permissions are not granted automatically and require runtime user approval for accessing resources.


181 questions inquired about SignatureOrSystem permissions. This is a restricted category of permissions, only available to system apps or apps signed with the same cryptographic key (indicating that both come from the same developer). Among these 181 questions \textit{Debugging}-related issues were highest with 87 questions, followed by \textit{Implementation Issues} (63).


\begin{tcolorbox}
\textit{\textbf{Summary RQ3:} Our analysis found that developers face problems mainly related to non-evolving SDK permissions. This indicates that official documentation may lack adequate detail, leading developers to seek additional help. A smaller number of questions concerned evolving permissions, and developers typically struggle with non-SDK permissions when dealing with evolution-related issues. Although these permissions are not frequently discussed, they are crucial to manage properly for security reasons. Alarmingly, most of the discussed permissions were either Dangerous or Signature level, which can pose high-security risks. The misuse of these permissions, especially in Android versions where they were not initially introduced, could lead to potential vulnerabilities. The Stack Overflow community often suggests workarounds for these permissions, which may suggest a lack of thorough understanding of proper permission usage. Therefore, developers must gain sufficient knowledge on permission use to maintain the security of their apps. Notably, the majority of the issues developers encountered related to Debugging and Implementation, especially with Dangerous permissions, which are not automatically granted and can cause issues at runtime. Fewer questions pertained to the SignatureOrSystem permissions, reserved for system apps or apps signed with the same key, further indicating challenges with more restricted permissions. In conclusion, correctly understanding and implementing permissions remains a substantial challenge for many developers, impacting the security of their applications.}
\end{tcolorbox}

\subsection{ Results of RQ4 : Alignment of Community Provide Information and the Official Documentation}

\textbf{Restriction Type Comparison with Androids' Official Information:} 

We have analyzed the restriction types of permissions discussed by developers with Android's official information. The summary of our analysis is presented in Table~\ref{table:restriction_comparison}. 22 questions inquired about 31 specific permissions. We aligned the permissions with the corresponding categorization present in the official documentation. Among these 31 restriction types, only 16 matched the official restriction lists.  


 Interestingly, we came across two instances where developers inquired about restriction types without mentioning any specific permissions, e.g., ``Is there any way to use blocked permissions?." In these instances, we did not have enough information to check the alignment with the documentation.

Among the analyzed answers, only 17 answers mentioned specific permissions, and all matched the official documentation. This is consistent with our observation that more experienced developers tend to provide answers.

\begin{table*}[t]
\centering
\scalebox{0.65}{
\begin{tabular}{lrrrrrrrr}
\hline
\multicolumn{1}{c}{\multirow{2}{*}{\begin{tabular}[c]{@{}c@{}}Restriction\\ Type\end{tabular}}} & \multicolumn{3}{c|}{\textbf{Questions}} & \multicolumn{3}{c}{\textbf{Answers}} \\ \cline{2-7} 
\multicolumn{1}{c}{} &
  \multicolumn{1}{c}{\begin{tabular}[c]{@{}c@{}}\# Mentioned\\ Restriction \\ Type\end{tabular}} &
  \multicolumn{1}{c}{\# Matched} &
  \multicolumn{1}{c}{\# Mismatched} &
  \multicolumn{1}{c}{\begin{tabular}[c]{@{}c@{}}\# Mentioned\\ Restriction\\ Type\end{tabular}} &
  \multicolumn{1}{c}{\# Matched} &
  \multicolumn{1}{c}{\# Mismatched}\\ \hline
SDK                                                                                             & 18     & 16    & 2   & 11   & 11    & -    \\
Blocklist                                                                                       & 4      & -     & 4      & 0     & -     & -     \\
Greylist-max-x                                                                                  & 4      & -     & 4     & -     & -     & -    \\
Unsupported                                                                                     & 3      & -     & 3     & -     & -     & -    \\\hline
Total                                                                                           & 31     & 16    & 12    & 11    & 11    & -     \\ \hline
\end{tabular}}
\caption{\label{table:restriction_comparison}Comparison between Community Provided Information and Official Documentation about Restriction Type}
\end{table*}


\textbf{Protection Level Comparison for SDKs with Android'sOfficial Information:}
We similarly analyzed whether the protection levels of permissions discussed by developers align with the official Android information. The summary of our analysis is presented in Table~\ref{table:Protection_comparison}. Only 126 questions contained 138 permissions with the corresponding protection level.

Our analysis showed that in 45 cases, the mentioned protection levels did not match the official documentation. There were 49 questions asking generic questions without indicating specific permissions, e.g., ``how to use dangerous permissions?" which prevented us from doing further analysis. 

Among the answers, 109 instances mentioned the protection level with the corresponding permissions. 85 (77.9\%) of them agreed with the official documentation.

\begin{table*}[t]
\centering
\scalebox{0.65}{
\begin{tabular}{lrrrrrrrr}
\hline
\multicolumn{1}{c}{\multirow{2}{*}{\begin{tabular}[c]{@{}c@{}}Protection\\ Level\end{tabular}}} & \multicolumn{3}{c|}{\textbf{Questions}} & \multicolumn{3}{c}{\textbf{Answers}} \\ \cline{2-7} 
\multicolumn{1}{c}{} &
  \multicolumn{1}{c}{\begin{tabular}[c]{@{}c@{}}\# Mentioned\\ Protection Level\end{tabular}} &
  \multicolumn{1}{c}{\# Matched} &
  \multicolumn{1}{c}{\# Mismatched} &
  \multicolumn{1}{c}{\begin{tabular}[c]{@{}c@{}}\# Mentioned\\ Protection Level\end{tabular}} &
  \multicolumn{1}{c}{\# Matched} &
  \multicolumn{1}{c}{\# Mismatched} \\ \hline
Dangerous                                                                                       & 60    & 42     & 18   & 62    & 47   & 15    \\
Normal                                                                                          & 41    & 29     & 12     & 32   & 28   & 4     \\
Signature                                                                                       & 37    & 22     & 15     & 15    & 10   & 5     \\\hline
Total                                                                                           & 138   & 93    & 45     & 109   & 85   & 29    \\ \hline
\end{tabular}}
\caption{\label{table:Protection_comparison} Comparison between Community Provided Information and Official Information about SDKs Protection Level}
\end{table*}




Furthermore, we analyzed a set of 1008 pairs of questions and their accepted answers. In this analysis, we found that while 871 questions discuss specific permissions, only 395 accepted answers do the same. There were 137 pairs where the question mentioned permissions, but the answers did not. Conversely, in 22 pairs, the questions did not mention permissions, yet the answers did. This suggests that an answer does not always discuss the same permissions as its associated question and vice versa. Both the question and accepted answer discuss permissions in 373 pairs only. We further investigated them. Our analysis revealed that while 83 unique SDKs were mentioned in the questions, only 65 were referenced in the answers. Six questions discussed five SDK permissions mirrored in their answers. Seven questions, each about a unique conditionally blocked permission, had their accepted answers mentioning the same. This could suggest developers are adding unnecessary SDK permissions in their apps; therefore, accepted answers do not discuss them. Interestingly, some developers provide accepted answers that mention \textit{Mismatched} permissions. In nine answers, they discussed four unique permissions that do not appear in the permission list for the specific version discussed in the answer. Therefore, there may be some permissions that are usable in specific versions but are not officially documented. Conversely, some permissions might have been removed from the documentation because they should not be used in certain versions but were not removed from the system properly. This could pose significant security risks for the apps.

\begin{tcolorbox}
\textit{\textbf{Summary RQ4:} Our findings suggest that developers rarely specify the restriction type or protection level of the permissions they are dealing with. When they do mention these aspects, they often do not align with what is stated in the official documentation. Interestingly, when a match does occur, it is more likely to be with the protection level than with the restriction type. This suggests that developers may not fully understand or be aware of the restriction types and protection levels. This lack of awareness could contribute to the incorrect use of permissions, potentially leading to the development of less secure apps. Additionally, there are some \textit{Mismatched} permissions in the accepted answers, which might cause potential security risks.}
\end{tcolorbox}

\section{Discussion \& Implications}
\subsection{Discussion}

Our study revealed that developers face a diverse range of challenges related to permissions. Our findings emphasize several important issues.

\begin{itemize}
\item \textbf{Most of the challenges are related to SDK Permissions:}  In other words, the majority of issues encountered by developers revolve around permissions that are officially documented. However, the significant number and diverse range of problems indicates that the official documentation might not offer sufficient information for developers to utilize the permissions correctly.\\



\item \textbf{Developers face their biggest challenges with  \textit{dangerous} permissions: } These permissions are not automatically granted to an application. Developers must request user approval during the application's runtime before the first use of the restricted functionality. In this case, incorrect use of permissions prevents apps from working correctly, raising developers' concerns.  \\


\item  \textbf{Misunderstanding of Permissions: }
We observed a small number of questions and answers that discussed permissions absent from the official documentation or the restriction lists of an indicated version of Android. 
The presence of these questions and answers is a clear indication of the developers' lack of knowledge.
This is not surprising given the numerous concerns expressed by developers about a lack of documentation and its confusing and contradictory nature, as our analysis showed.

We also observed that developers often fail to specify the restriction type and protection level of the permissions in their questions or provide incorrect information about these aspects. This shows that developers are unaware of the restriction lists published by Google or use them incorrectly. 

Such oversights could lead to misuse of permissions. Therefore, it is vital to offer recommendation support that can aid developers in dealing with permission-related challenges and ensure the correct usage of Android permissions.

\item  \textbf{Developer faces challenges due to Android evolution: }
We observed that developers seem unaware of the dependencies related to permission or the required Android API version necessary for specific permissions. Although Android provides limited updates to address these device-specific issues, many problems persist. These issues are often linked to requirements and compatibility problems due to updates. Even though tools are available (e.g., ~\cite{mahmudAndroid2021}) to identify these compatibility problems, developers continue to face challenges. The fast-evolving nature of the permission model, coupled with a lack of documentation, requires more thorough support than what is currently provided by Google.\\

\item \textbf{Lack of community-provided solutions: } Our study also discovered that many questions about permission-related challenges remain unanswered. Only 67.1\% of such questions find solutions, and just 30.2\% receive accepted answers. This is comparatively low compared with the overall question-answer rate of 91.3\% (\cite{bhat2014min}) and the accepted answer rate of 53.0\% (\cite{yazdaninia2021characterization}). We also observed an interesting trend: developers are not getting the exact types of solutions they seek for their challenges (Table ~\ref{table:mapping}). This mismatch could be why the rate of accepted answers is lower than average on Stack Overflow. This emphasizes the need for more targeted, relevant solutions to developers' questions regarding Android permissions. Therefore, it underscores the necessity for more precise and relevant responses to developers' questions regarding Android permissions. It also points to the need for a system that facilitates the exploration of permission-related challenges and proposes solutions.\\


\item \textbf{Developers point up the need for up-to-date and reliable official documentation: }Generally, documentation has been a long-standing challenge in software development (\cite{wedyan2020impact}). Android development is no exception. We observed that Google’s documentation sources sometimes provide conflicting
information about permissions. This inconsistency could lead to the
inappropriate use of permissions, exposing data and phone resources to unnecessary risk. 
We found that more than half of the errors are directly associated with permissions. For example,  developers do not use required permissions, data-sensitive permissions without justification, or improperly implement runtime permissions. This highlights a need for better documentation of the Android platform and its permissions system. Previous studies corroborate our findings (\cite{felt2011android},~\cite{sellwood2013sleeping}), stating that while Android provides developer documentation, the information specifically about permissions is limited. This limited information can result in errors due to the lack of reliable permission guidance. The analysis results highlighted the importance of providing clear and comprehensive documentation and guidelines for correctly implementing permissions-related features.\\


\item \textbf{Less experienced developer faces more challenges: } Based on our observations, it appears that developers who ask questions on Stack Overflow generally have lower reputation scores compared to those who provide answers. This aligns with our analysis, suggesting that, given the lack of sufficient documentation, less experienced developers rely on Stack Overflow as a primary source for permission handling advice. On the other hand, even when faced with inadequate documentation, more experienced developers are better prepared to handle such challenges and offer valuable guidance.
\end{itemize}

\subsection{Implications}

The findings from this study carry several implications for both Android app developers and the broader mobile app development community.

\begin{itemize}
\item \textbf{Implications for Developers: }
 Our analysis shows that developers often strive to create more secure, user-friendly, and compliant applications, although they have challenges adapting to the dynamic nature of the Android permissions system. The presence of mismatched permissions in both questions and answers gives perhaps one of the most actionable lessons for developers. These mismatches are likely to occur due to a lack of knowledge and, therefore, can be resolved through careful analysis of restriction lists and available documentation. \\




\item \textbf{Implications for Researchers: } The provided insights into developers'  challenges with Android permissions could facilitate further research in this domain. We make the collected data available to the community. Examining the inquiries can uncover gaps, while exploring the effectiveness and accuracy of solutions can, for example, help determine best practices. This dual strategy of investigating the problems and their solutions provides a holistic perspective, which could greatly aid current and future research endeavors.\\
    
\item \textbf{Implications for Google: }
Our study sheds light on issues Android app developers often face with app permissions. As the Android platform continues to evolve, so does its permissions system, making it challenging for developers to keep up-to-date with the latest developments in the presence of inadequate documentation. Understanding the challenges associated with app permissions in the evolving Android ecosystem can lead to more informed decision-making and understanding of the existing gaps in the documentation. For example, the integration of restriction lists with the official documentation  could create a comprehensive  one-stop resource for Android permissions.\\

\end{itemize}



\section{Threats to Validity}

In this section, we discuss the threats to the validity of this study and how
these threats were mitigated in our research. \\

\textbf{Internal Validity }refers to the extent to which the observed findings can be attributed to the ``treatment" itself rather than other extraneous factors. We used search tags, such as ``android," ``permission," and ``android-permission," to identify the related posts in Stack Overflow, and this is a threat to the internal validity of our study because we might have missed other tags. To mitigate this threat, we used the list of permissions and matched it with the title and body (except code) of Stack Overflow questions. Therefore, any post discussing permissions can be added to our dataset. Then we manually verified and classified our dataset. This verification stage might pose a risk to our study, as the manual validation of the classification could have been subject to human mistakes and bias. To lessen this risk, two authors categorized the datasets. Any disagreements among the authors were settled through conversation until consensus was achieved.
\\

\textbf{External validity} is regarding the scope of generalizability of our findings. Our study relies solely on Stack Overflow data, which may not be representative of all Android developers or the broader Android ecosystem. Therefore, our findings may not be generalized to other populations or platforms. Our research could be further enhanced by including more sources (i.e., Google Issue Tracker)to mitigate this threat. \\

\textbf{Conclusion Validity } pertains to the relationship between the outcomes and the dataset in our study. We gathered our dataset of Android permission-related questions based on specific tags and permission names found in the question title and body. However, there could be more Android permission-related questions that are not identified by these tags and names. Some developers might pose questions without realizing that the problem is related to permissions, and answers may still pertain to permissions. This issue could affect the total number of extracted Android permission-related questions, consequently influencing the results and conclusions of our research. Furthermore, the manual coding method used in the study may include the authors' personal judgment. Even with a joint review and revision by the two authors, biases or inconsistencies might still exist, which could potentially impact the study's conclusions. \\


\section{Conclusion}
This research aimed to investigate the challenges developers encounter while using Android permissions, the solutions provided by the community, the specific restriction types and protection levels that cause the most issues, and the accuracy of information shared within the developer community about permissions.
Through this study, we found that developers commonly face issues with debugging, especially related to the misuse or incorrect use of Android permissions. Issues also frequently arose due to the changing landscape of Android permission policies, suggesting developers struggle to adapt to the evolving system. Despite the importance of correctly implementing permissions to prevent misuse and compatibility issues, our analysis showed that the response rate and quality of answers to these issues within the community were insufficient.
Most community responses provided implementation guidelines, yet they often did not match the developers' specific challenges. This mismatch could contribute to the low acceptance rate of provided solutions. Further support mechanisms, like improved documentation or additional tools, could enhance developers' understanding and correct use of permissions.
Developers faced most problems with non-evolving SDK permissions and those of the Dangerous protection level, which are not automatically granted and require runtime access. There were also instances where discussed permissions did not exist in the stated Android version, further highlighting the knowledge gap in this area.
Finally, it became evident that developers often lack a complete understanding of the permissions' restriction types and protection levels. This knowledge deficiency may lead to incorrect permission usage, creating potential security risks.

In light of our research, future research can be done on developing supportive tools to guide developers in accurately implementing permissions. These tools could assist with compatibility issues, offer real-time alerts about changes in the permission landscape, and suggest solutions to common problems. A parallel area of study is investigating the user's comprehension of permissions, as they often need to grant permissions during runtime. Furthermore, a longitudinal study tracking the evolution of challenges and corresponding solutions in line with the progression of the Android permission system could provide valuable insights. Such a study would ensure that the understanding of Android permissions matures alongside the platform's evolution. 

In conclusion, this research underscores the necessity of addressing the identified challenges related to Android permissions. Developers' struggles indicate the need for improved community support, better educational resources, and the development of tools to understand and correctly apply permissions. This could result in safer applications, a more informed developer base, and enhanced user privacy and security.
\\
\\

\begin{footnotesize}
\begin{longtable}{|l|p{2.9in}|}

\hline
Category &  Example \\ \hline
\endfirsthead
\endhead
Documentation - Related & \\
\hspace{0.1in} 1. Lack of Documentation & 
``Android 13 - READ\_EXTERNAL\_STORAGE \_PERMISSION still  usable (...)I didn't  find any  information about this in the official documentation, but to be  honest the documentation is  focusing on media files only  without any word about other file types(...)" \\ 
&\\
\hspace{0.1in}2. Confusing Documentation &
``I came to know the privacy  changes for Android 10 and  I'm quite clear that third-party  apps won't be able to get  IMEI now. But one thing from  documentation is creating confusion for me.(...) which means that on Android devices with API LEVEL 28 or lower, this method returns null or placeholder data, even if the app is having READ\_PHONE\_STATE permission. Right? (...) any idea about it? or am I  misinterpreting it?" \\ 
&\\   
\hspace{0.1in}\begin{tabular}[c]{@{}l@{}} 3. Contradictory\\ Documentation\end{tabular} &
 ``(...)So I want to setNotificationVisiblity  to VISIBLITY\_HIDDEN. According to docs I need to set the permission in  AndroidManifest.xml. Where is  the permission? I can't find permission android.permission. DOWNLOAD\_WITHOUT \_NOTIFICATION in  Manifest.permission. Reference: \url{https://developer.android.com/reference/android/Manifest.permission} (...)"  \\ \arrayrulecolor{black}\hline

Problems with Dependencies & \\
\hspace{0.1in}1. Version Dependency &
``Permission is only not working  on Version 8.0.0 \& Version 8.1.0. Rest all other Versions including Android Version 9, permission is working like a charm" \\\arrayrulecolor{blue}\hline

\hspace{0.1in}2. Device Dependency &
``(...)It's working on other devices I have tested but not working in Xiaomi's MIUI devices. If possible please provide me with a work around. I desperately needed this permission as my app want to change ringtones.\\ 
 &\\ 
\hspace{0.1in} \begin{tabular}[c]{@{}l@{}}3. Version \& Device \\Dependency\end{tabular} &
``I'm facing screen overlay issue on Android 8.0. Mentioned below are the permission which are mentioned in my manifest.xml file - (...)I've tested the app in Android 7.0 \& it is working fine on it. I'm getting this issue on Android 8.0 devices like Nexus 6P / Google Pixel devices.\\ \arrayrulecolor{black}\hline

Debugging & \\
\hspace{0.1in} 1. Error Handling &\\
\hspace{0.2in} \begin{tabular}[c]{@{}l@{}}a.Usage of Data Sensitive \\Permission\end{tabular} &\\
\hspace{0.3in}
\begin{tabular}[c]{@{}l@{}}i. Usage of Permission \\Without Core \\Functionality\end{tabular} & ``I added a description before taking permission from the user but still getting the warning. Your app is requesting the following permission which is used by less than 1\% of functionally similar apps: WRITE\_CALL\_LOG. Users prefer apps that request fewer permissions and requesting unnecessary permissions can affect your app's visibility on he Play Store" \\ &\\  
\hspace{0.3in}
\begin{tabular}[c]{@{}l@{}}ii. User Data Policy \\  Declaration\end{tabular} &``I got the following error: Your APK or Android App Bundle is using permissions that require a privacy policy: (android.permission.READ\_PHONE \_STATE). This is an unexpected behaviour right? Why will a vanilla app require this permission? How can I solve this issue?" \\ &\\
\hspace{0.2in}\begin{tabular}[c]{@{}l@{}}b. Absence of Required \\ Permission\end{tabular} &
``I am building an Android app via android studio. One major feature of the app is sending SMS but I cannot seem to get it to run correctly on different devices due to permission errors.(...)I also just got this error hen sending message java.lang.SecurityException: Neither user 10205 nor current process has android.permission.READ\_PHONE \_STATE" \\ &\\   \arrayrulecolor{blue}\hline

\hspace{0.2in}\begin{tabular}[c]{@{}l@{}}c. Exception or Warning \end{tabular} &
``I want a uninstall an app without poping the ``Do you want to uninstall this app?" dialog. I have tried using(...) Getting the below error (...) Exception occurred while dumping: java.lang.NullPointerException(...) Note: INTERACT\_ACROSS\_USERS \_FULL permission is added in the AndroidManifest file and also the app is signed as a system app" \\ &\\ 

\hspace{0.1in} \begin{tabular}[c]{@{}l@{}}2. Functional Improvement\end{tabular} &
``If the user denies the permission the first time a dialog should appear where the use of the phone number is explained (...) To check if the user has ticked never ask again, I use the method shouldShowRequestPermissionRationale. But the method always returns false, even if I never ticked never ask again" \\ \arrayrulecolor{black}\hline

Conceptual Understanding &\\
\hspace{0.1in} 1. Rationale Analysis & ``An application I work on calls TelephonyManger.getCallState() to find out if the call state is CALL\_STATE\_IDLE. (...) For any alternatives it suggests - you must either have READ\_PHONE\_STATE or carrier privileges (...) Why was it okay for 12 years and suddenly not?"
\\&\\&\\&\\&\\  
\hspace{0.1in}
\begin{tabular}[c]{@{}l@{}}2. Understanding\\ Permission Usage\end{tabular} &
``Do we require any permission for storing images above android Q or R and Media API would be good for storing images? As google have changed storage security in android 11 and WRITE\_EXTERANL\_STORAGE won't work. So can we save images on phone without permission of user?"\\ \arrayrulecolor{black}\hline \hline

Implementation Issues &\\
  \hspace{0.1in}1. Functionality Implementation &\\ 
  \hspace{0.2in}a. Feature specific &\\
  \hspace{0.3in}i. Location Services & ``Android Google Map - How to access user's location if permissions are granted(...) have the corresponding permissions added in the Manifest file. Now my question is, if allPermissionsGranted is true, how can I access the user's location as a LatLng object."\\ &\\  \arrayrulecolor{blue}\hline

  \hspace{0.3in}\begin{tabular}[c]{@{}l@{}}ii. Data \& File \\ Management\end{tabular} & ``I try to save files on my smartphone. Since I have no SD card installed I must save the data to my internal memory(...). I also tried the following permissions:READ\_EXTERNAL\_STORAGE WRITE\_EXTERNAL\_STORAGE (...) Does anyone know how to access these folders?"\\ &\\
  \hspace{0.3in}\begin{tabular}[c]{@{}l@{}}iii. Services \& Background\\ Processes\end{tabular} & ``I am trying to run as a background task this code(...) The task of it is to grab data from an online database and display it to a toast (...)Is it possible to have this be executed as a service?" \\ &\\
  \hspace{0.3in}\begin{tabular}[c]{@{}l@{}}iv. Bluetooth \& Wifi\\ Connectivity\end{tabular} & ``i want that when i click on the battery saver button then the ``WIFI" and ``BLUETOOTH" should be turned off. So, can anybody help me for the code of this that when I click on the button then i can turn off the ``WIFI" and ``BLUETOOTH" both ?" \\ &\\  
  \hspace{0.3in}v. UI \& Navigation & ``how to display VOIP call screen above Always On  Display screen? (...). if Always On Screen (AOD) is turn on I can hear the ringtone but I can't see the screen with the buttons for an answer or decline the call (...) and permissions: ``android.permission.WAKE\_LOCK" ``android.permission.DISABLE \_KEYGUARD" Thank you all!''\\ &\\ 
  \hspace{0.3in}vi. Camera \& Media Access &``(...)The app is capturing image perfect on wrong password, but the issue is when the app is in a kill or background state, then the camera is not capturing any image. (...)I think that's because of dangerous permission, but there are many apps available on PlayStore with these features.Any suggestions will be helpful" \\  &\\ 
 
  \hspace{0.3in}\begin{tabular}[c]{@{}l@{}}vii. Device \& Hardware\\ Interaction\end{tabular} & ``I tried working with the SensorManager but (...) problematic proximity sensors.(...) Is there anyway to listen to the PROXIMITY\_SCREEN\_OFF\_WAKE\_LOCK activity so I could know when to change the audio output source?"\\  \arrayrulecolor{blue}\hline

 \hspace{0.2in}
  \begin{tabular}[c]{@{}l@{}}b. Permissions Management\end{tabular} &
  ``How to get user typed url from browser using PACKAGE\_USAGE\_STATS permission? I want to get the browser url (From all browser that is which browser using their mobile) which is typed by user on their browser. I can get that URL by using AccessibilityService. I just want to know the same how to get the url using PACKAGE\_USAGE\_STATS permission in android." \\ &\\
\hspace{0.1in}2. Alternative Solution  &\\
  \hspace{0.2in}\begin{tabular}[c]{@{}l@{}}a. Adapting Permission\\ Policy \& Changes\end{tabular} & ``(...)We found that we could use the permission BIND\_DEVICE\_ADMIN to make the process unkillable. However it will be deprecated soon and we have no idea how long it will remain functional.(...) Does anybody have a workaround/solution or does somebody know how much longer they will support the BIND\_DEVICE\_ADMIN permission?" \\ &\\ 
  \hspace{0.2in}b. Improved Solution & ``Read/write to Downloads folder deniedafter targeting Android SDK version 29 (...)So, my question is: what is the right, official, robust way to ensure access to the Download folder for a Flutter app, in the upcoming Android version?(...)" \\ &\\  
  
   \hspace{0.2in}\begin{tabular}[c]{@{}l@{}}c. Avoiding \\ Permission Usage\end{tabular} & ``(...) Is there a way to ``download" files without any permission. Open them, getting the input from the file and work with it (...)"\\ &\\ 
   \hspace{0.2in}d. Power Consumption & ``Fetch continuous location changes with less battery drainage (...) So I want to know what should be done to use the minimum battery power and fetch the continuous location changes."\\ \arrayrulecolor{black}\hline\hline
\caption{\label{table:example}The Examples of Developers' Questions for Each Challenge Category (*Blue colored line refers to the continuation of the table)}

\end{longtable}
\end{footnotesize}


\balance
\bibliographystyle{elsarticle-harv}
\bibliography{bibliography.bib}
\end{document}